\documentclass[fleqn,usenatbib]{mnras}

\usepackage{newtxtext,newtxmath}
\usepackage[T1]{fontenc}

\DeclareRobustCommand{\VAN}[3]{#2}
\let\VANthebibliography\thebibliography
\def\thebibliography{\DeclareRobustCommand{\VAN}[3]{##3}\VANthebibliography}

\usepackage{graphicx}	% Including figure files
\usepackage{xcolor}
\usepackage{hyperref}
\newcommand{\sizeFigM}{0.7\columnwidth}
\newcommand{\sizeFigL}{1.0\columnwidth}
\newcommand{\orcid}[1]{\href{https://orcid.org/#1}
{\includegraphics[width=10pt]{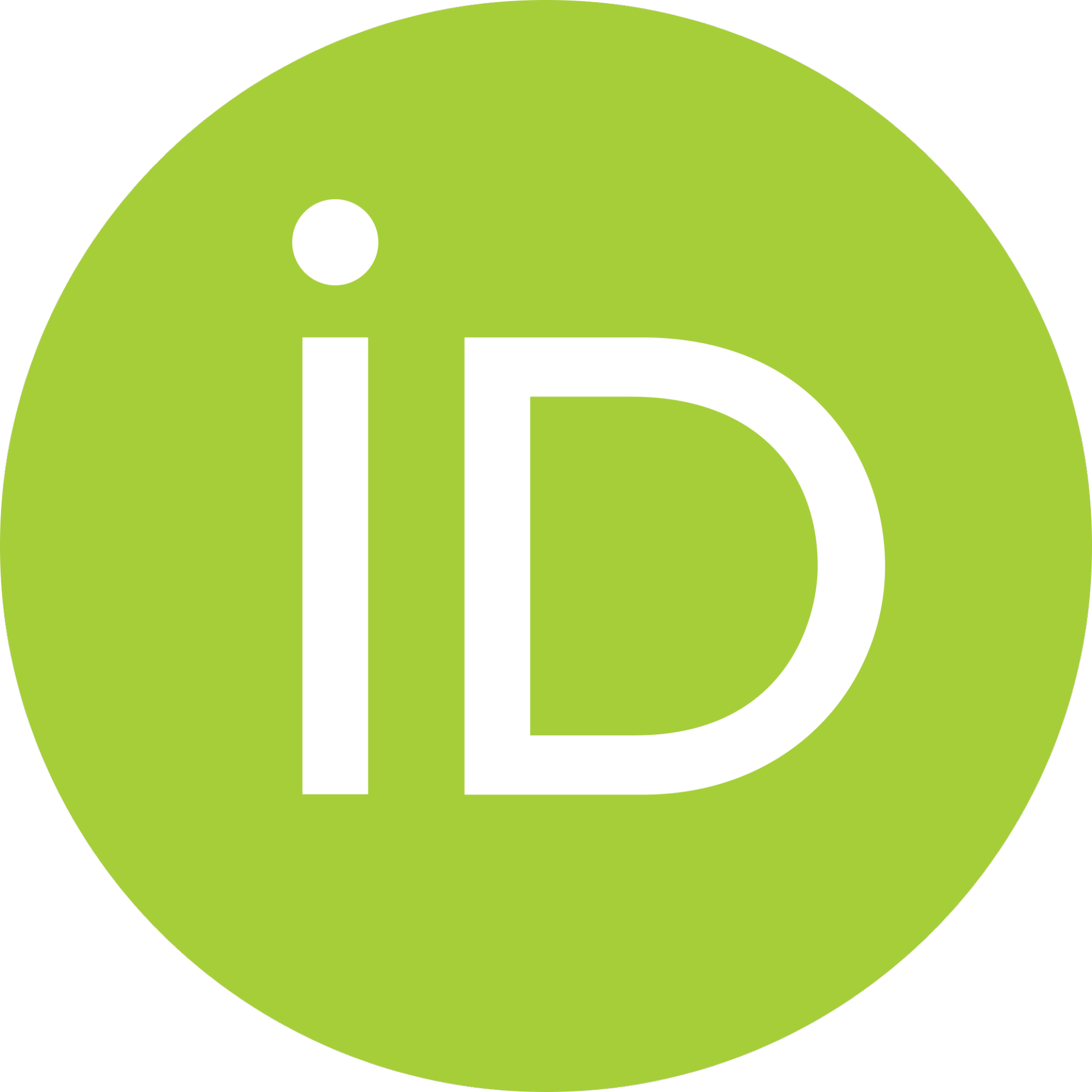}}}
\title[High-Energy Neutrinos in Galaxy Clusters]{ High-Energy Neutrino Production in Clusters of Galaxies}
 \author[S. Hussain et al.]{
Saqib Hussain, $^{\orcid{0000-0002-0458-0490}~ ^{1}}$ \thanks{E-mail: s.hussain2907@gmail.com (SH)}
Rafael {Alves Batista}, $^{\orcid{0000-0003-2656-064X}~ ^{2}}$ \thanks{E-mail: r.batista@astro.ru.nl}
Elisabete M. {de Gouveia Dal Pino}, $^{\orcid{0000-0001-8058-4752}~ ^{3}}$ \thanks{E-mail: dalpino@iag.usp.br}
\newauthor
 and Klaus Dolag, $^{\orcid{0000-0003-1750-286X}~ ^{4,5}}$ %\thanks{kdolag@mpa-garching.mpg.de}
 \\
$^{1,3}$  Institute of Astronomy, Geophysics and Atmospheric Sciences (IAG), University of S\~ao Paulo (USP), S\~ao Paulo, Brazil \\
$^{2}$ Radboud University Nijmegen, Department of Astrophysics/IMAPP, 6500 GL Nijmegen, The Netherlands\\
$^{4}$University Observatory Munich, Scheinerstr. 1, 81679 Munchen, Germay\\
$^{5}$Max Planck Institute for Astrophysics, Karl-Schwarzschild-Str 1, 85741 Garching, Germany
 }
\date{Accepted XXX. Received YYY; in original form ZZZ}
\pubyear{2020}
\begin{document}
\label{firstpage}
\pagerange{\pageref{firstpage}--\pageref{lastpage}}
\maketitle

\begin{abstract}
Clusters of galaxies can potentially produce cosmic rays (CRs) up to very-high energies via large-scale shocks and turbulent acceleration. 
Due to their unique magnetic-field configuration, CRs with energy $\leq 10^{17}$ eV can be trapped within 
these structures
over cosmological time scales, and generate secondary particles, including neutrinos and gamma rays, through interactions with the background gas and photons. In this work we compute the contribution from clusters of galaxies to the diffuse neutrino background. We employ three-dimensional cosmological magnetohydrodynamical simulations of structure formation to model the turbulent intergalactic medium. We use the distribution of clusters within this cosmological volume to extract the properties of this population, including mass, magnetic field, temperature, and density.
We propagate CRs in this environment using multi-dimensional  Monte Carlo simulations across different  redshifts (from $z \sim 5$ to  $z =0$), considering all  relevant photohadronic, photonuclear, and hadronuclear interaction processes. 
We find that, for CRs injected with a spectral index $\alpha = 1.5 - 2.7$ and cutoff energy $E_\text{max} = 10^{16} - 5\times10^{17} \; \text{eV}$, clusters contribute to a sizeable fraction to the diffuse flux observed by the IceCube Neutrino Observatory, but most of the contribution comes from clusters with $M \gtrsim 10^{14} \; M_{\odot}$ and redshift $ z \lesssim 0.3$.  If we include the cosmological evolution of the CR sources, this flux can be even higher.

\end{abstract}


\section{Introduction}

The IceCube Neutrino Observatory reported evidence of an isotropic distribution of neutrinos with $\sim$~PeV energies \citep{aartsen2017contribution, aartsen2020time}. Their origin is not known yet, but the isotropy of the distribution suggests that they are predominantly of extragalactic origin. They might come from various types of sources, such as galaxy clusters~\citep{murase2013testing, hussain2019propagation}, starbursts galaxies,
galaxy mergers, AGNs~\citep{murase2013testing, kashiyama2014galaxy, anchordoqui2014icecube, khiali_dalpino2016,fang2018linking}, supernova remnants~\citep{chakraborty2015diffuse, senno2015extragalactic}, gamma-ray bursts~\citep{hummer2012neutrino, liu2013diffuse}. Since neutrinos can reach the Earth without being deflected by magnetic fields or attenuated due to any sort of interaction, they can help to unveil the sources of ultra-high-energy cosmic rays (UHECRs) that produce them.

Their origin and that of the diffuse gamma-ray emission  are among the major mysteries in astroparticle physics. The fact that the observed energy fluxes of UHECRs, high-energy neutrinos, and gamma rays are all comparable suggests that these messengers may have some connection with each other \citep{ahlers2018opening, alves2019open, ackermann2019astrophysics}. 
The three fluxes could, in principle, be explained by a single class of sources  \citep{fang2018linking}, like starburst galaxies or galaxy clusters \citep[e.g.,][for reviews ]{murase2008cosmic,kotera2009propagation,alves2019open}.

Clusters of galaxies form in the universe possibly through violent processes, like accretion and  merging of smaller structures into larger ones \citep{voit2005tracing}. These processes release large amounts of energy, of the order of the gravitational binding energy of the clusters ($\sim 10^{61} - 10^{64}~ \text{erg}$).
Part of this energy is depleted  via shock waves and turbulence through the intracluster medium (ICM), which accelerate CRs to relativistic energies. These can be also re-accelerated by similar processes in more diffuse regions of the ICM, including relics, halos,  filaments, and  cluster mergers  \citep[e.g.,][for reviews]{brunetti2014cosmic, brunetti2020second}.
Furthermore, clusters of galaxies are attractive candidates for UHECR production due to their extended sizes ($\simeq$~Mpc) and suitable magnetic field strength ($\sim 1 \; \mu\text{G}$) \citep[e.g.,][]{fang2018linking, kim2019filaments}. 
Those with energies $E > 7\times 10^{18}$ eV have most likely an extragalactic origin  \citep[e.g.,][]{aab2018large, batista2019cosmogenic}, and those with $E \lesssim 10^{17}$ eV  are believed to have Galactic origin \citep[see e.g., ][]{blasi2013origin, amato2018cosmic}, although the exact transition between galactic and
extragalactic CRs is not clear yet \citep[see e. g.,][]{aloisio2012transition, parizot2014cosmic, giacinti2015escape,thoudam2016cosmic, kachelriess2019transition}.

CRs with  $E \lesssim 10^{17}$ eV can be confined within clusters for a time comparable to the age of the universe 
\citep[e.g.][]{hussain2019propagation}. 
This confinement 
makes clusters efficient sites for the production of secondary particles including, electron-positron pairs,  neutrinos and gamma rays due to their interaction with the 
%local gas, 
thermal protons and photon fields \citep[e.g.][]{berezinsky1997clusters,rordorf2004diffusion, kotera2009propagation}. Non-thermal radio to gamma-ray and neutrino observations are, therefore, the most direct ways of constraining the properties of CRs in clusters~\citep{berezinsky1997clusters,wolfe2008broadband,yoast2013winds,zandanel2015high}.
Conversely, the diffuse flux of gamma rays and neutrinos depend on the energy budget of CR protons in the ICM. Clusters also  naturally can introduce a spectral softening due to the fast escape of high-energy CRs from the magnetized environment which might explain the second knee that appears around $\sim 10^{17}$~eV, in the CR spectrum~\citep{apel2013kascade}.

To calculate the fluxes of CRs and secondary particles from clusters, there are many analytical and semi-analytical works~\citep{berezinsky1997clusters,wolfe2008broadband,murase2013testing}, but in most of the approaches, the ICM model is overly simplified by assuming, for instance, uniform magnetic field and gas distribution. 
There  are more realistic numerical approaches in \citet{rordorf2004diffusion} and \citet{kotera2009propagation}  exploring the three-dimensional (3D)  magnetic fields of clusters.
More recently, \citet{fang2016high} estimated the flux of neutrinos from these objects assuming an injected CR spectrum $\propto E^{-1.5}$, an isothermal gas distribution, a radial profile for the total matter (baryonic and dark) density profile, and a Kolmogorov turbulent magnetic field with coherence length $\sim 100 \; \text{kpc}$. They found these estimates to be comparable to IceCube measurements.
Here we revisit these analyses by employing a more rigorous numerical approach. We take into account the non-uniformity of the gas density and magnetic field distributions in clusters, as obtained from MHD simulations. We consider additional factors such as the location of CR sources within a given cluster, and the obvious mass dependence of the physical properties of clusters. This last consideration is important because massive clusters ($\gtrsim 10^{15} \; M_{\odot}$) are much less common than lower-mass ones ($\lesssim 10^{13} \; M_{\odot}$). Consequently, clusters that can confine CRs of energy above PeV for longer are probably more relevant for detection of high-energy neutrinos.

Our main goal is to derive  the contribution of clusters to the diffuse flux of high-energy neutrinos. To this end, we follow the propagation and cascading of CRs and their by-products in the   cosmological background simulations by \citet{dolag2005constrained}. We use the Monte Carlo code CRPropa~\citep{batista2016crpropa} that accounts for all relevant photohadronic, photonuclear, and hadronuclear interaction processes. Ultimately, we obtain the CR and neutrino fluxes that emerge from the clusters.

This paper is organized as follows: in section~\ref{sec:NumMethod} we describe the numerical setup for both the cosmological background simulations and for CR propagation through this environment; in section~\ref{sec:results} we 
% present our results of the 3D-MHD simulations, flux
characterize the 3D-MHD simulations and present our results for the fluxes
of CRs  and neutrinos; in section~\ref{sec:Discussion} we discuss our results; finally,  in section~\ref{sec:conclusion} we draw our conclusions.

%%%%%%%%FIGURE
\section{Numerical Method}\label{sec:NumMethod}
\subsection{Background MHD Simulation}\label{sec:MHD-Sim}

To study the propagation of CRs in the ICM we consider the large scale cosmological 3D-MHD simulations performed by \citet{dolag2005constrained}, who employed  the Lagrangian smoothed particle hydrodynamics (SPH) code GADGET~\citep{springel2001gadget,springel2005cosmological}.
These simulations  capture the essential features of 
the mass, temperature, density, and 
magnetic field distributions in galaxy clusters, filaments and voids.

\begin{figure}
\centering
\includegraphics[width=\columnwidth]{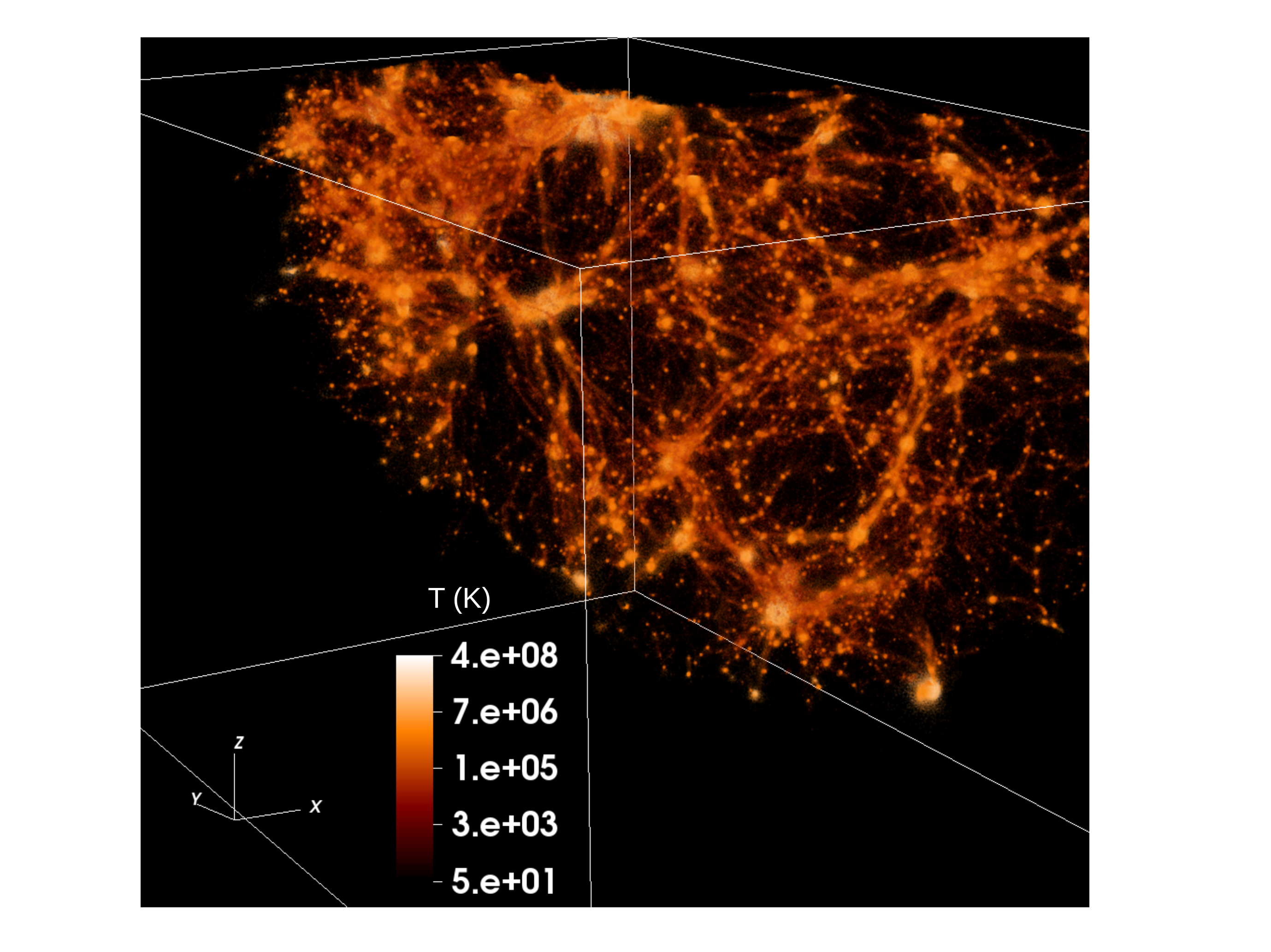}
\includegraphics[width=\columnwidth]{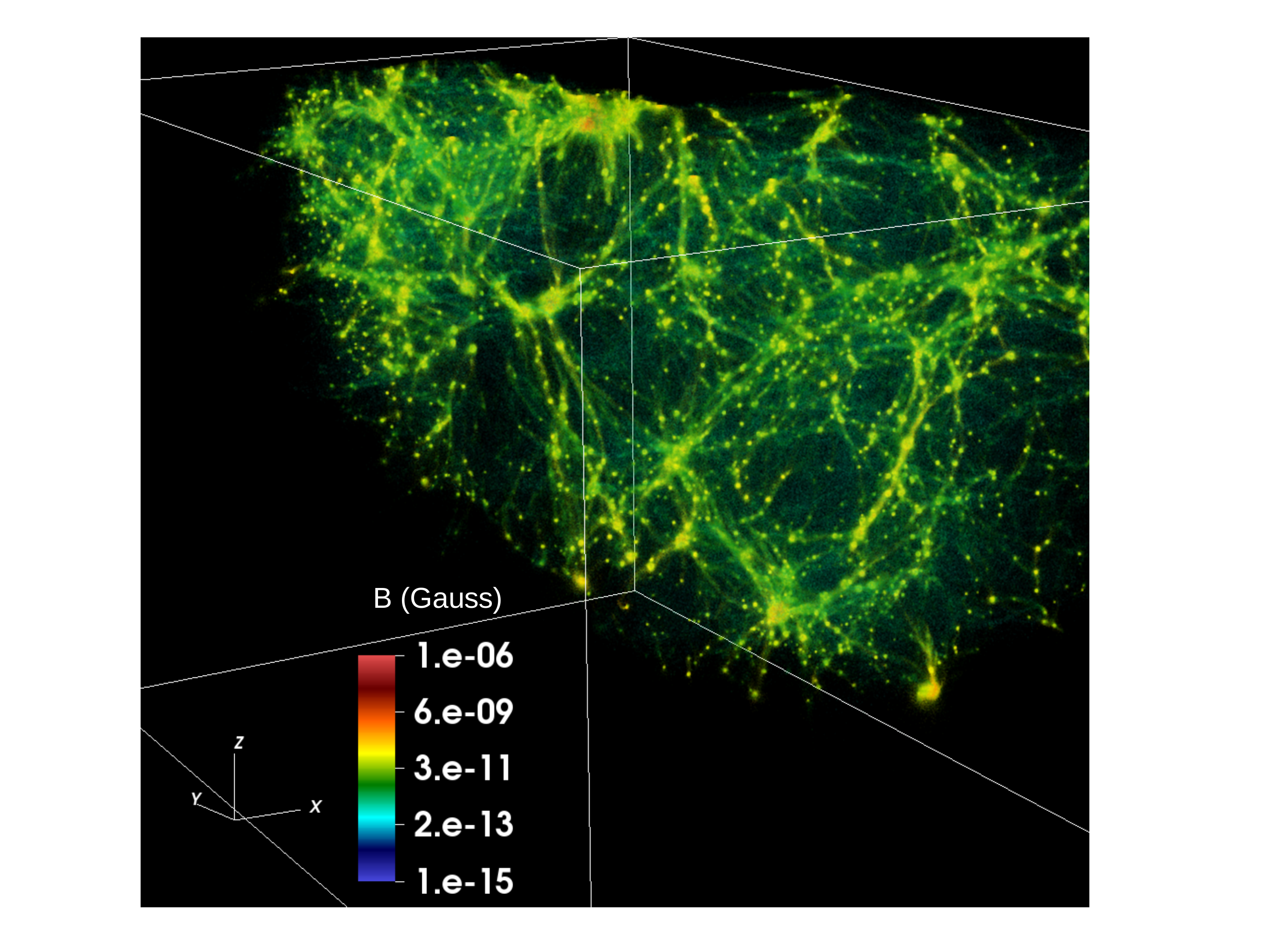}
\caption{This figure shows the temperature (upper panel) and magnetic field (lower panel) for one of the eight regions of our  background 3D-MHD cosmological simulation at redshift $z = 0.01$, with dimension $240~ \text{Mpc}^3$,  performed by \citet{dolag2005constrained}.}
\label{fig:filaments-Tmp}
\end{figure}

We  consider here seven snapshots of these simulations with redshifts $z = 0.01; \; 0.05; \; 0.2; \; 0.5; \; 0.9; \; 1.5; \; 5.0$, each having the same volume $(240 \; \text{Mpc})^{3}$. We have divided the domain of each snapshot into eight regions. Fig.~\ref{fig:filaments-Tmp} shows the temperature and magnetic-field distributions for one of the regions, at redshift $z= 0.01$.

The filaments in Fig.~\ref{fig:filaments-Tmp} are populated with galaxy clusters and have dimensions $\sim 50 \; \text{Mpc}^3$,  while the voids have dimensions of the same order, which are compatible with observations \citep[e.g.,][]{govoni2019radio, gouin2020probing}. 
In this simulation, the comoving intensity of the seed magnetic field was chosen to be $B = 2\times10^{-12} \; \text{G}$, which leads to a quite reasonable match with the field strength observed in different clusters of galaxies today. Feedback and star formation were not included in these cosmological simulations. The background cosmological parameters  assumed 
are $h \equiv H_{0}/(100 \; \text{km~s}^{-1} ~\text{Mpc}^{-1}) = 0.7$, $\Omega_m = 0.3$, $\Omega_{\Lambda} =0.7$, and the baryonic fraction $\Omega_b/\Omega_m = 14\; \%$.

%%%%%%%%%%%%%%%%%%%%%%%%%%%%%%%%%%
\subsection{Simulation Setup for Cosmic Rays}\label{sec:Sim-Setup}

The  simulations described in the previous section provide the background magnetic field, gas density and temperature distributions of the ICM. 
In order to study the CR propagation in this environment,  
we employ the CRPropa 3 code \citep{batista2016crpropa}, with stochastic differential equations~\citep{merten2017crpropa}.

In these simulations, we assume that CRs are composed only by protons. We consider all relevant interactions during their propagation including photohadronic, photonuclear, and hadronuclear processes, namely photopion production, photodisintegration, nuclear decay,  proton-proton (pp) interactions, and adiabatic losses due to the expansion of the universe. The cosmic microwave background radiation (CMB)  and the extragalactic background light (EBL) are two essential ingredients, but other contributions comes from the hot gas component of the ICM, of temperatures between $\sim 10^6 - 10^8$~K, that produces bremsstrahlung radiation \citep{rybicki2008radiative} and serves as target for pp-interactions. This  is calculated in  section~\ref{sec:cosmological-background}.

\subsubsection{Cosmic-ray Propagation}\label{sec:CR-Prop}

To investigate the flux of different particle species and the change of their energy spectrum,
we use the Parker transport equation, which is a simplified version of the Fokker-Planck equation. It gives a good description of the transport of CRs for an isotropic distribution in the diffuse regime. It is given by:
\begin{equation}\label{b1}
\frac{\partial n}{\partial t} + \vec{u}.\nabla n = \nabla .(\hat{\kappa}\nabla n)+ \frac{1}{p^2}\frac{\partial}{\partial p} 
\left(p^2 \kappa_{pp} \frac{\partial n}{\partial p}\right)+\frac{1}{3}(\nabla  \vec{u})\frac{\partial n}{\partial \ln p} + S(\vec{x},p,t).
\end{equation}
Here $\vec{u}$ is the advection speed, $\hat{\kappa}$ is the spatial diffusion tensor, $p$ is the absolute momentum, $\kappa_{pp}$ is the diffusion coefficient of momentum used to describe the reacceleration, n is the particle density, $\vec{x}$ gives position and $S(\vec{x}, p, t)$ is the source of CRs (distribution of CRs at the source).

Propagation of CRs can be diffusive or semi-diffusive, depending on the Larmor radius ($r_\text{L} = 1.08 E_{15}/B_{\mu \text{G}}$ pc)  of the particles and the magnetic field of the ICM. The diffusive regime corresponds to $r_\text{L} \ll R_\text{cluster}$, and the semi-diffusive is for $r_\text{L} \gtrsim R_\text{cluster}$, wherein $R_\text{cluster}$ is the radius of the cluster, typically $\sim 1 \; \text{Mpc}$. Because $B \sim \mu\text{G}$, for the energy range of interest ($10^{14} - 10^{19} \; \text{eV}$), $r_\text{L} \ll R_\text{cluster}$, so we are in the diffusive regime.
CRs in this energy range would be confined completely by the magnetic field of the clusters for a time longer than the Hubble time ($t_\text{H} \sim 14$ Gyr) \citep[e.g.][]{fang2018linking}.
For instance, a CR with energy $\sim10^{17}$ eV in a cluster of mass $\sim 10^{14}\; M_{\odot}$ with central magnetic field strength $\sim 10^{-6} \; \mu\text{G} $ has $r_\text{L} \sim 0.1$ kpc much smaller than the size of the cluster ($\sim 2$ Mpc) and the trajectory length of this CRs inside the cluster is $\sim 10^{3}$ Mpc. The  confinement time for this CR can be calculated as $t_{\text{con}} \sim 1000~ \text{Mpc} / c \; \sim t_\text{H}$  
\citep[e.g.][]{hussain2019propagation}. 
Hence, CRs with energy $E>10^{17} \; \text{eV}$ have more chances to escape the magnetized cluster environment. 
The flux of CRs that can escape a cluster depends on its mass and magnetic-field profile, with the latter  directly correlated with the density distribution, being larger in denser regions.

\section{Results}\label{sec:results}

\subsection{Cosmological Background}
\label{sec:cosmological-background}

\begin{figure*}
\centering
\includegraphics[width=0.32\textwidth]{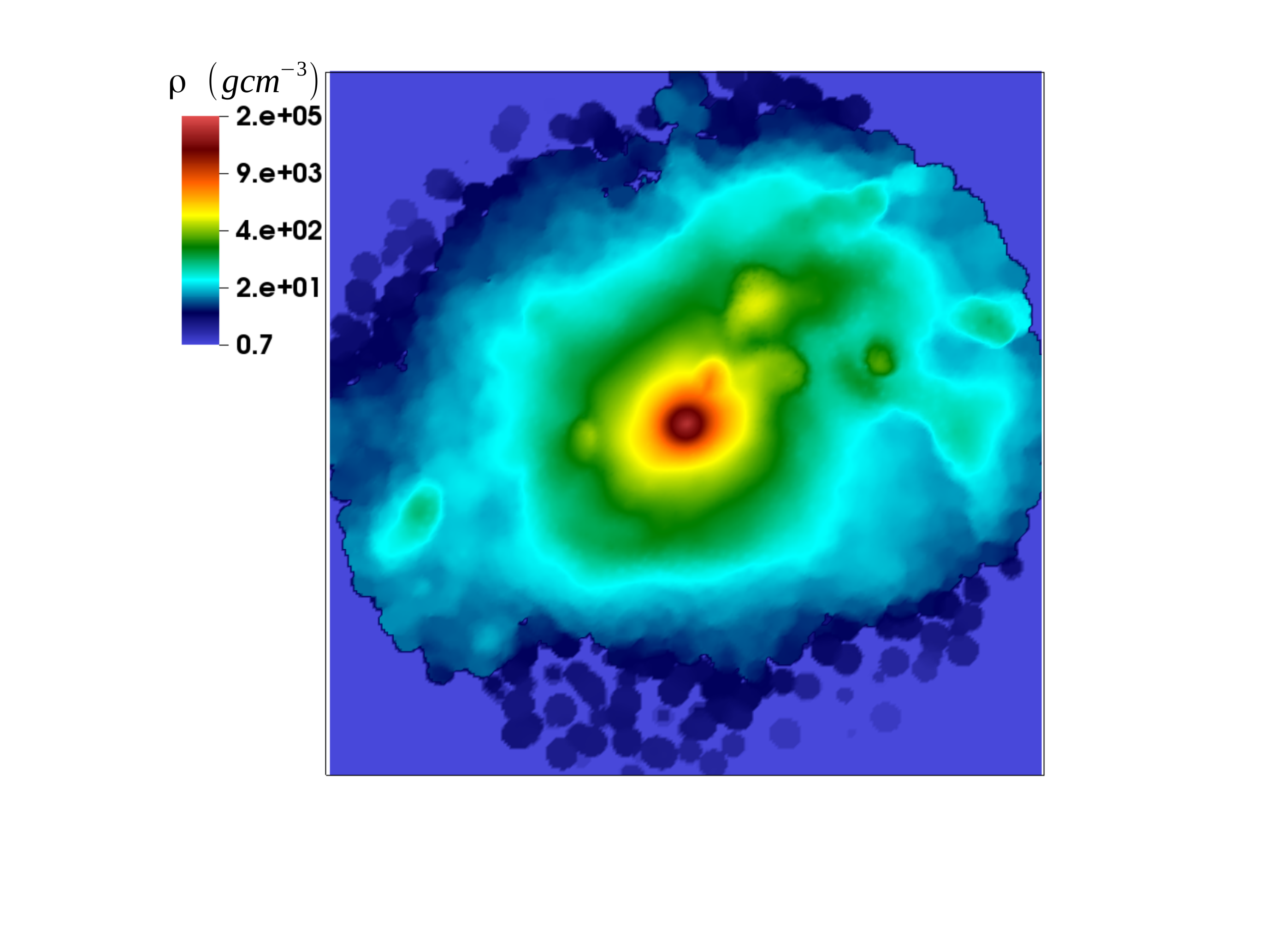} 
\includegraphics[width=0.32\textwidth]{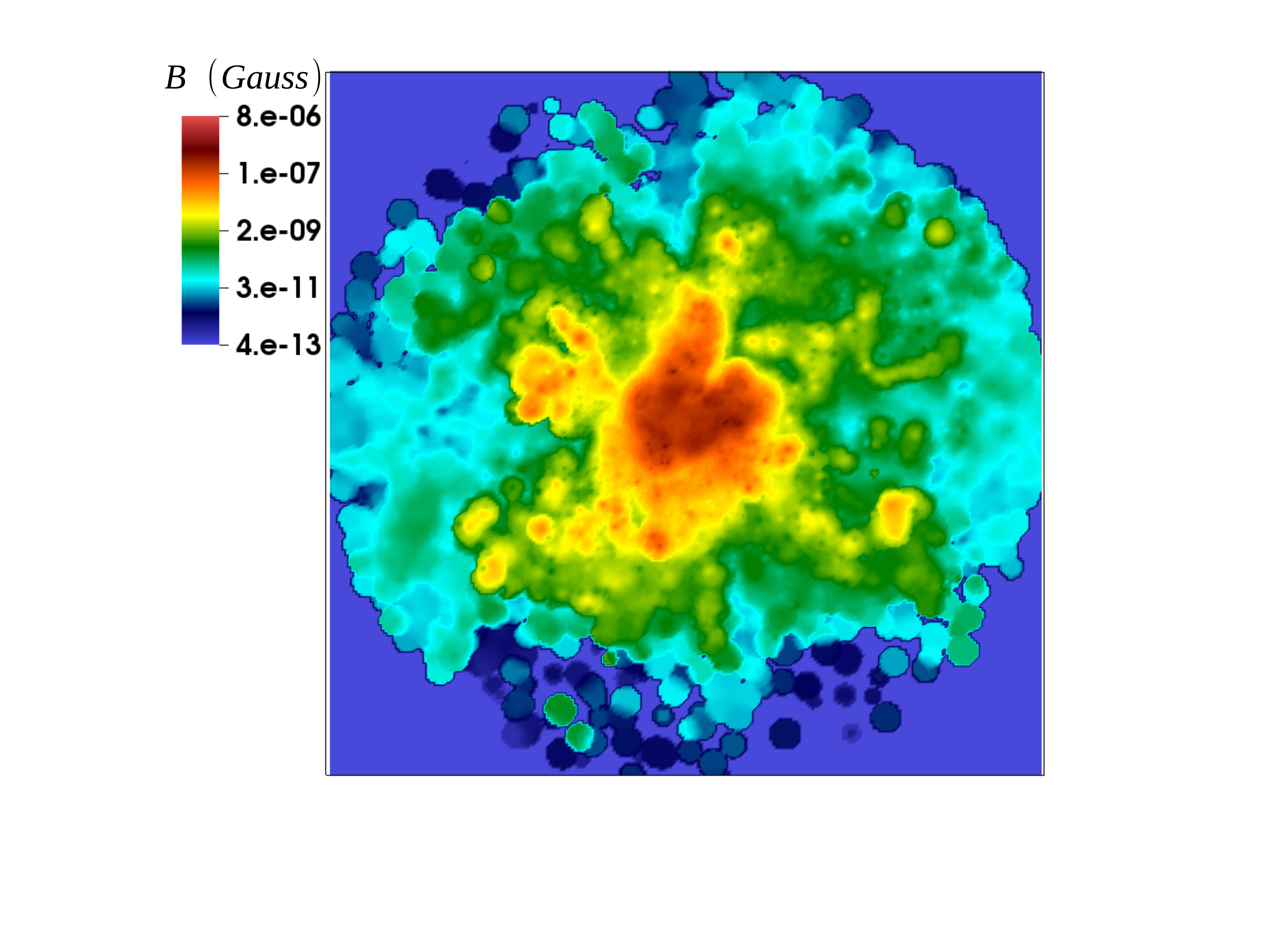} 
\includegraphics[width=0.32\textwidth]{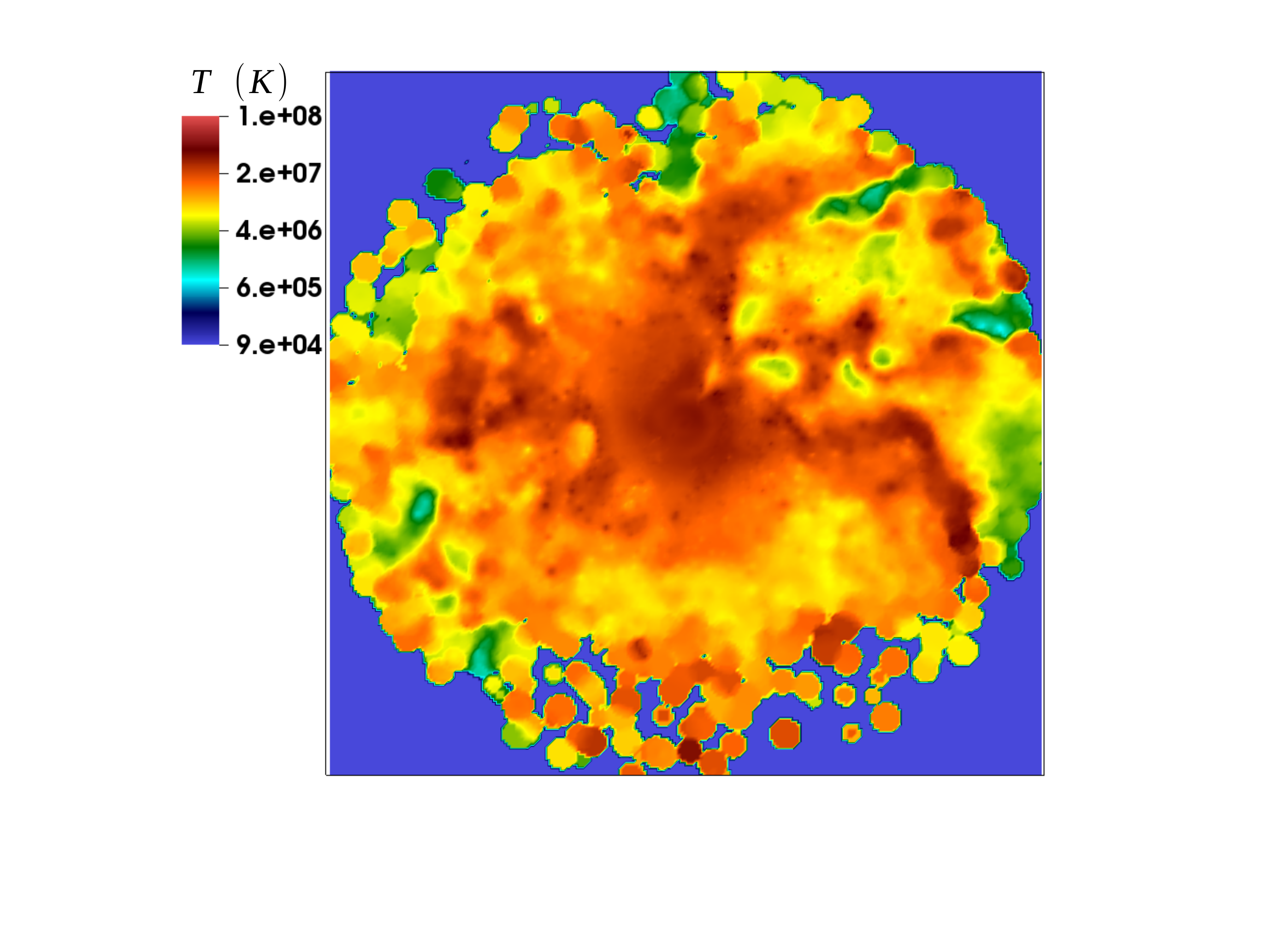} 
\includegraphics[width=0.32\textwidth]{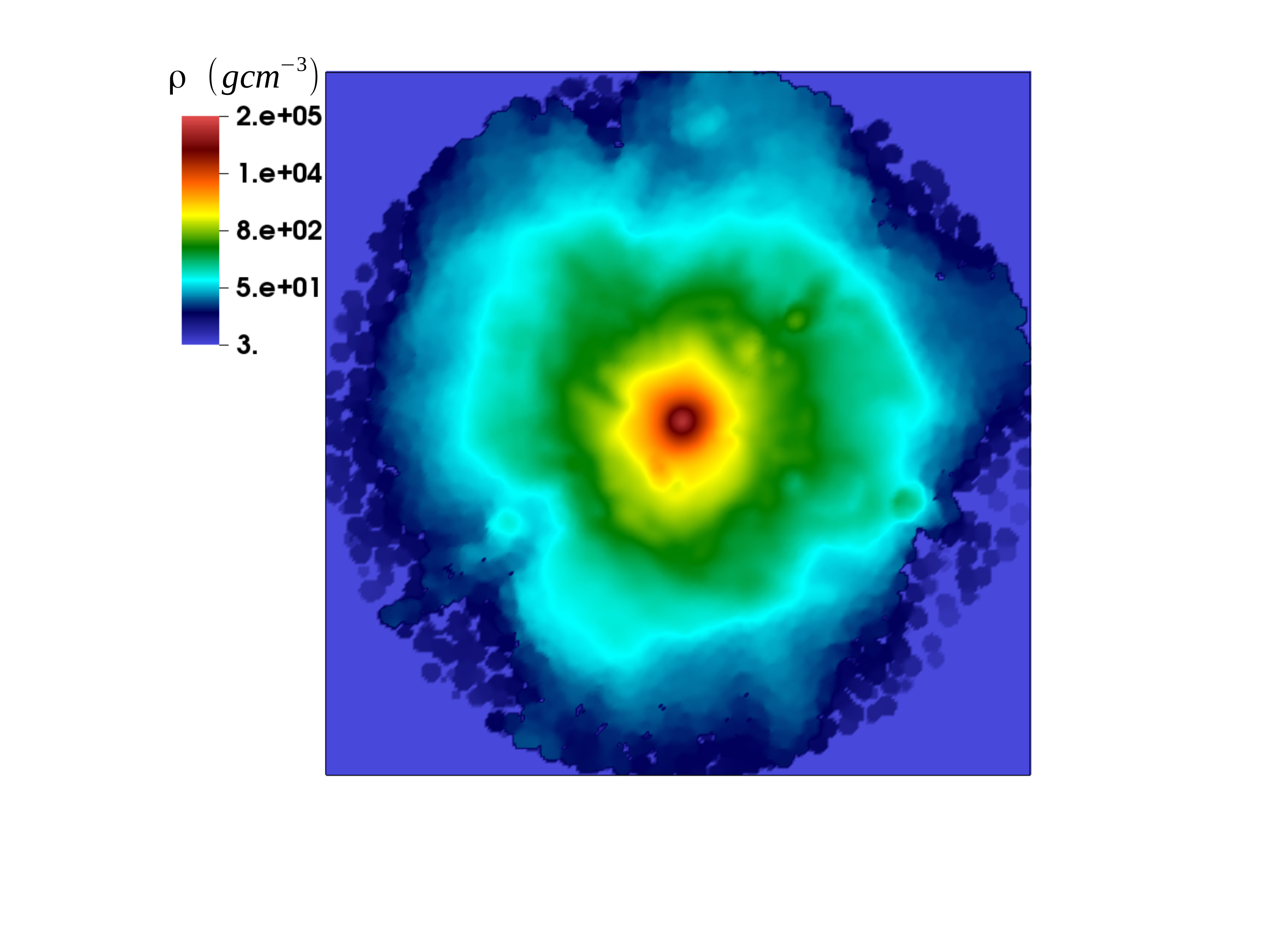} 
\includegraphics[width=0.32\textwidth]{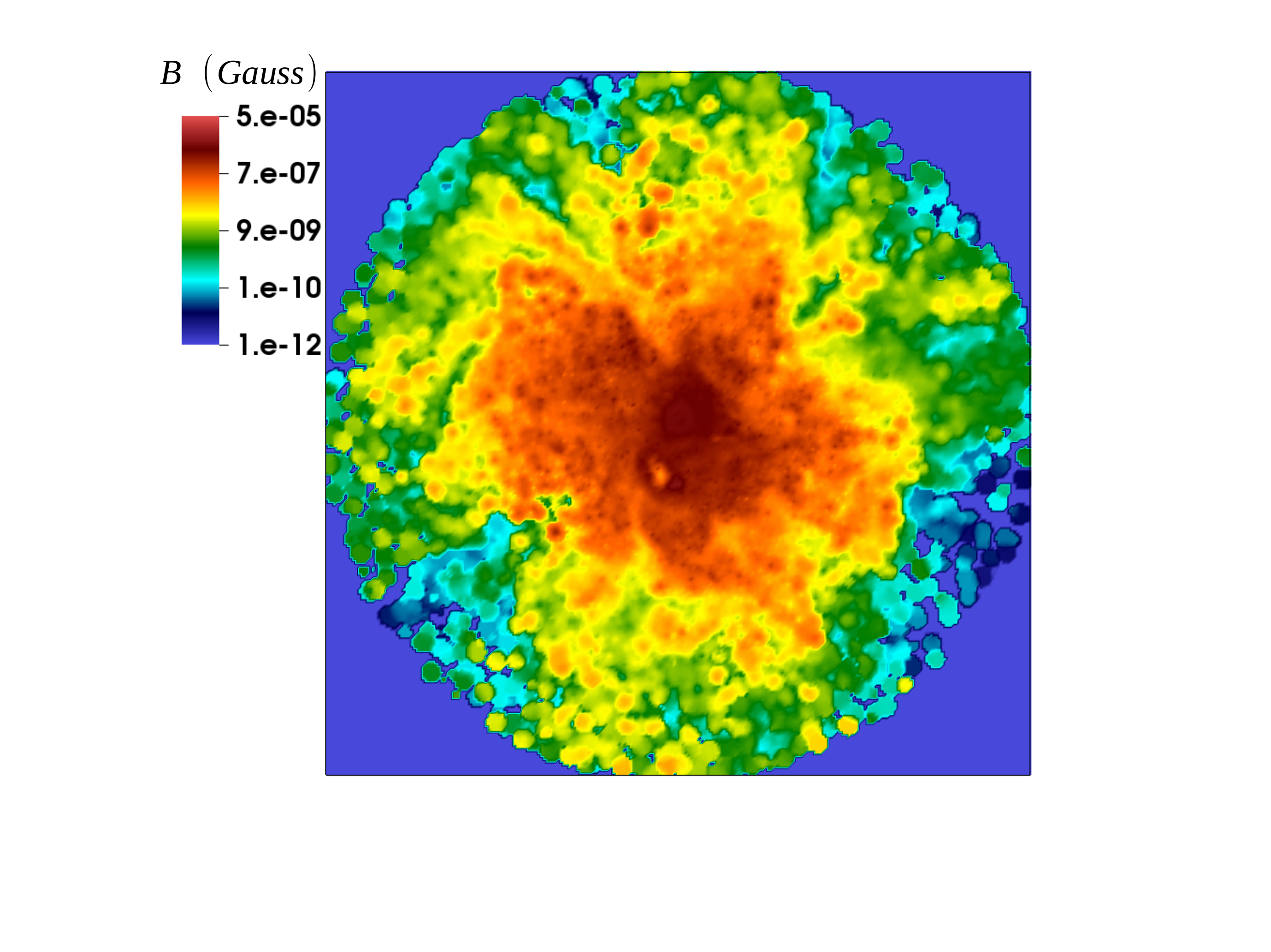} 
\includegraphics[width=0.32\textwidth]{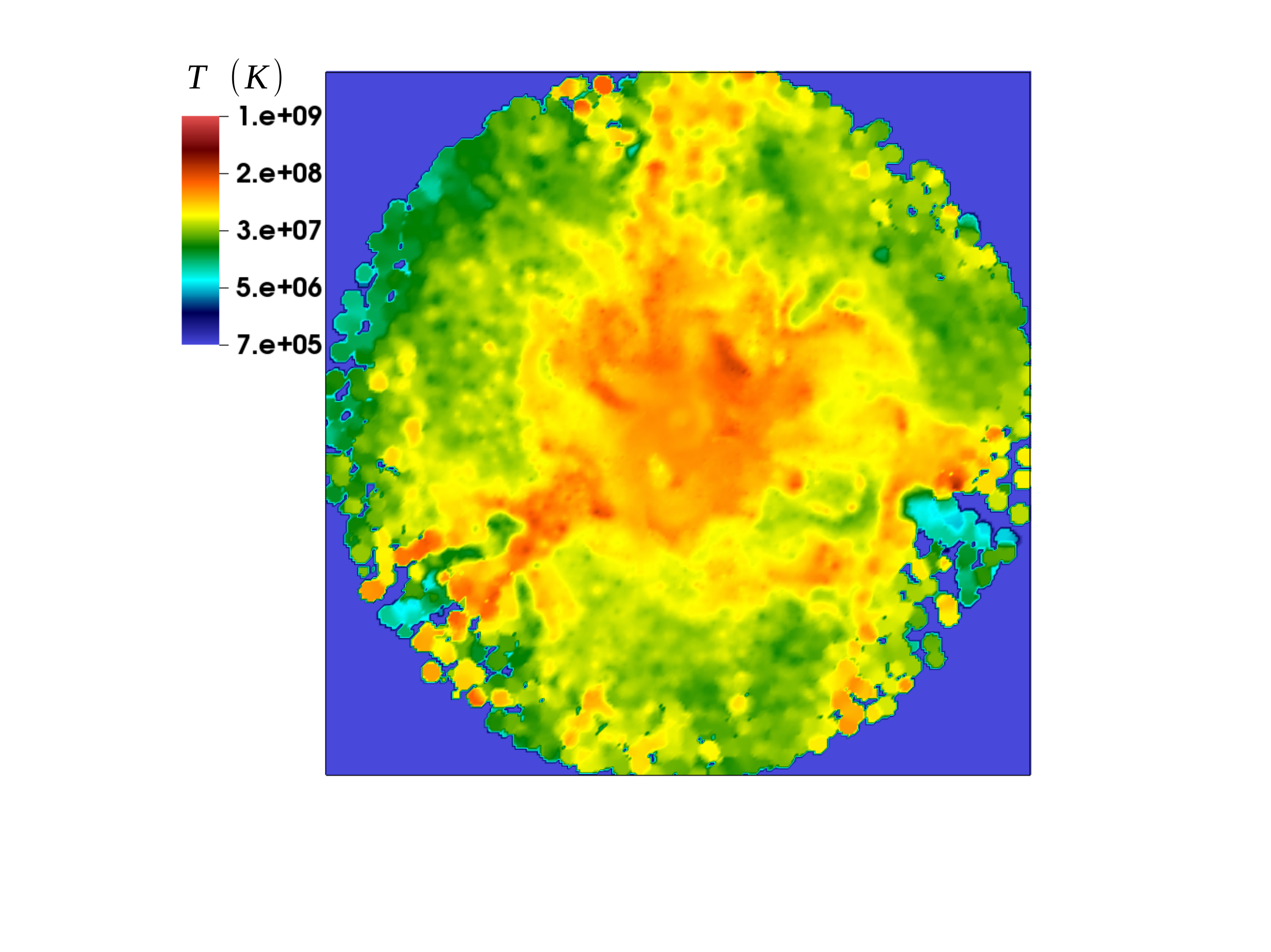} 
\caption{Maps of gas density (left column), magnetic field
(middle column), temperature (right column) of two clusters of masses $\sim 10^{14} M_{\odot}$ (upper panels) and  $\sim 10^{15} M_{\odot}$ (bottom panels), at redshift $z=0.01$. }
\label{fig:Cluster-contour}
\end{figure*}

Our background simulation includes seven snapshots in the redshift range $0.01 < z < 5.0$. 
We have identified clusters in the densest regions of the isocontour maps of the whole volume, in  each snapshot (see Fig.~\ref{fig:filaments-Tmp}).
 We then selected five clusters
 with distinct masses ranging from $10^{12}$ to $10^{16} \; M_{\odot}$, which we assumed to be 
 representative of all the clusters in the corresponding snapshot. Finally, we injected CRs in each of these
 clusters to study their  propagation and production of  secondary particles. As an example, Fig.~\ref{fig:Cluster-contour} illustrates relevant properties for two of these clusters with masses $\sim 10^{14} \; M_{\odot}$ (left panel) and $\sim 10^{15} \; M_{\odot}$ (right panel) at redshift $z=0.01$.
 To estimate the total mass of a cluster from the simulations, we integrated the baryonic and dark matter densities within a volume of $2$ Mpc, assuming an approximate spherical volume. We note that  this specific evaluation is not much affected by the deviations from spherical symmetry that we detect in Fig. \ref{fig:Cluster-contour}.
%for simplicity

To illustrate general average properties of the simulated clusters, we converted the  Cartesian into  spherical coordinates and divided the cluster in $10$ concentric spherical  shells of different radii ($R_{\text{shell}}$). Starting from the center of the cluster, the shells were first divided in intervals of $100 \; \text{kpc}$, then between $300 \; \text{kpc}$ and $1500 \; \text{kpc}$, they were divided in intervals of $200 \; \text{kpc}$, and the last shell in the outskirts was taken between $1500 \; \text{kpc} < r < 2000 \; \text{kpc}$.

\begin{figure}
\centering
\includegraphics[width=\sizeFigL]{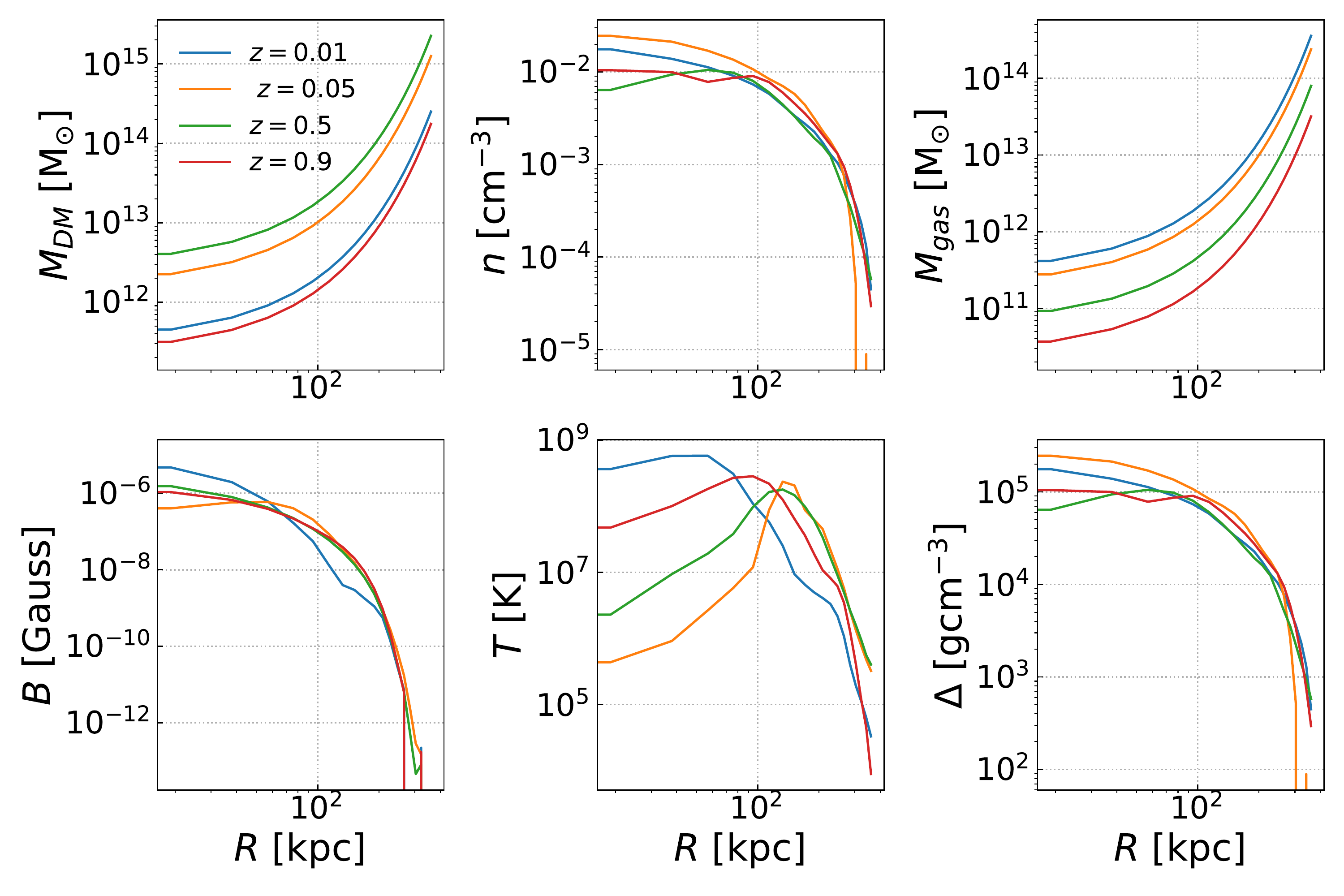}
\caption{ 
Volume-averaged profiles as a function of the radial distance from the center for a cluster of mass $M \sim 10^{15} \; M_{\odot}$, at  four different redshifts. The quantities shown are: dark-matter mass (top left); gas number density (top center); gas mass (top right); magnetic field (bottom left); temperature (bottom-center) and  overdensity (bottom right). 
} \label{fig:ClusterProp}
\end{figure}

\begin{figure}
\includegraphics[width=\sizeFigL]{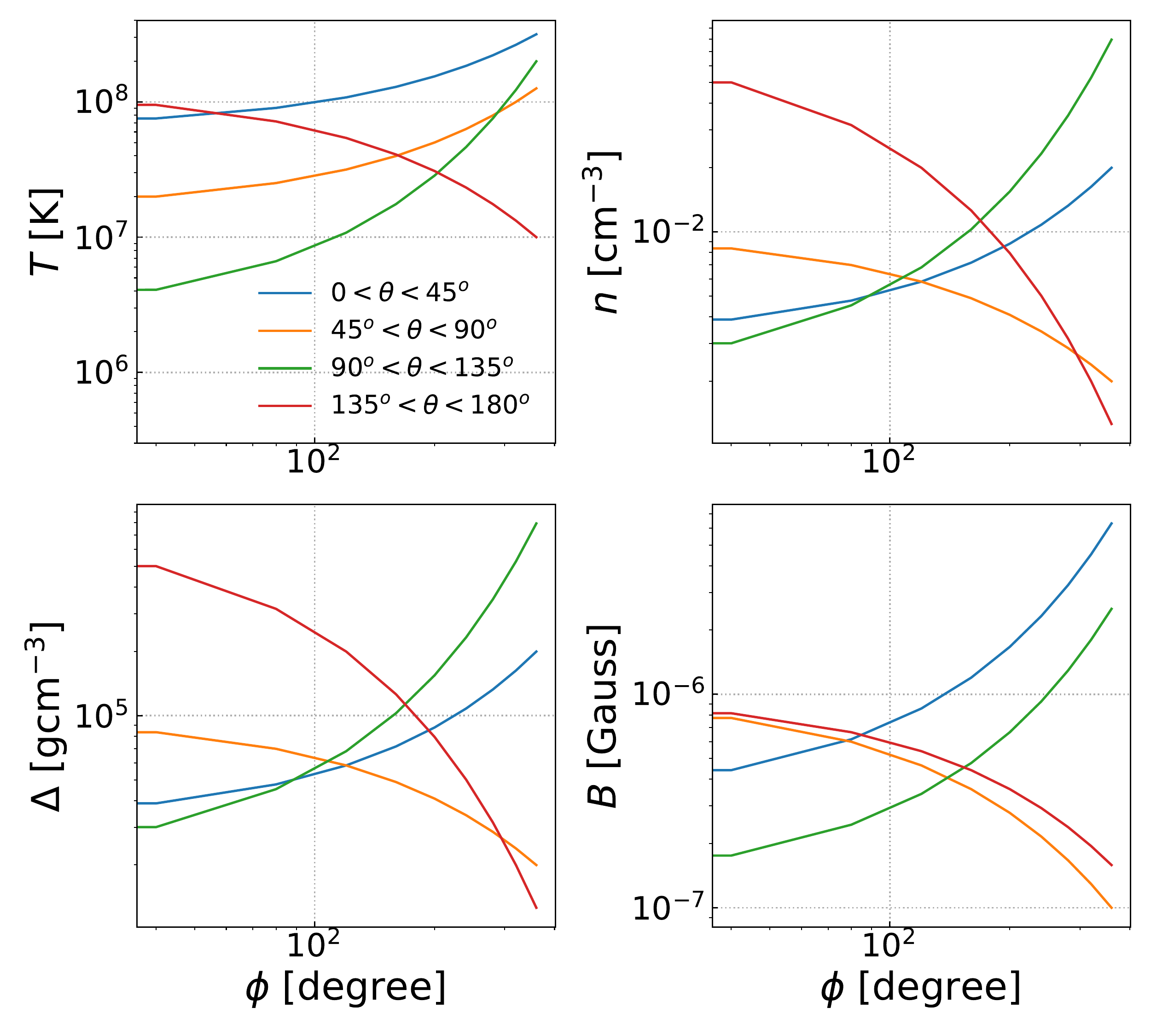}
\caption{
Volume-averaged profiles as a function of the azimuthal ($\phi$) angle for different latitudes ($\theta$), within  a radial distance $R= 300$~kpc from the center, for a  cluster of mass $M \sim 10^{15} \; M_{\odot}$. From top left to bottom right clockwise,  temperature, gas number density, overdensity and  magnetic field.}
\label{fig:Cluster_angular}
\end{figure}

Fig.~\ref{fig:ClusterProp} depicts volume-averaged profiles of different quantities as a function of the radial distance for a cluster of mass $\sim 10^{15} \; M_{\odot}$ at four different redshifts. The  overdensity in Fig.~\ref{fig:ClusterProp} (bottom-right panel) is defined as $\Delta = \rho(r) / \rho_{\text{bary}}$, where $\rho(r)$ is the total density at a given point and $\rho_{\text{bary}}$ is the mean baryonic density,   $\rho_{\text{bary}} = \Omega_{\text{bary}} \times \rho_{\text{crit}}$, $\rho_{\text{crit}} = 3 H^2 / 8\pi \text{G}$.  We see that, in general, these radial profiles are very similar across the cosmological time, except for the temperature that varies non-linearly with time by about four orders of magnitude in the inner regions of the cluster.
Fig.~\ref{fig:Cluster_angular} shows profiles for the temperature, gas density, magnetic field and overdensity for a cluster of mass $\sim 10^{15} \; M_{\odot}$,  as a function of the  azimuthal ($\phi$) angle for different latitudes ($\theta$), within a radial distance of $R= 300 \; \text{kpc}$, at a redshift $z = 0.01$. We see that there are substantial variations in the angular distributions of all the quantities. These variations characterize a deviation from spherical symmetry that may affect the emission pattern of the CRs and consequently secondary gamma rays and neutrinos.

\begin{figure}
\centering
\includegraphics[width=\sizeFigM]{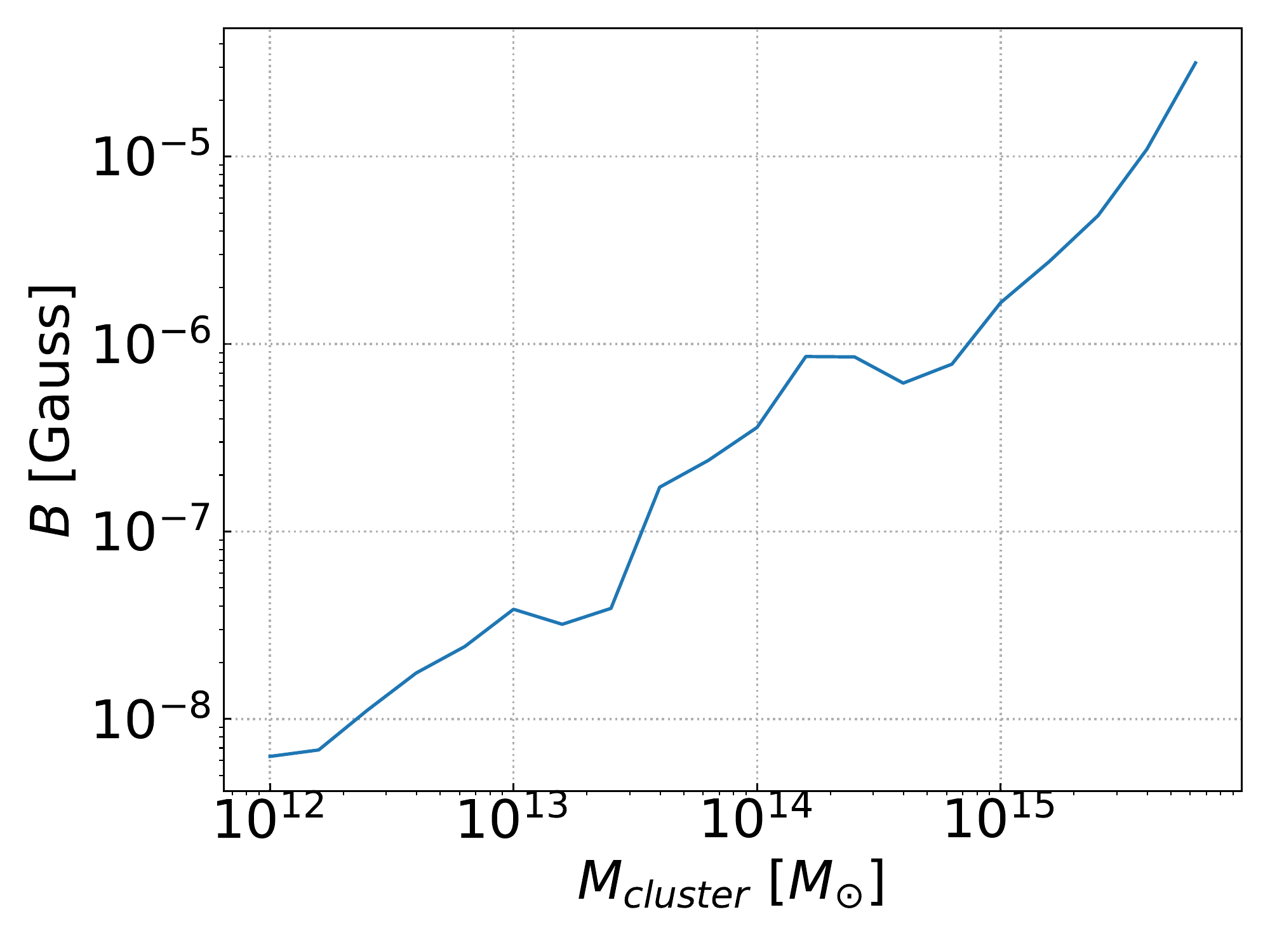}
\includegraphics[width=\sizeFigM]{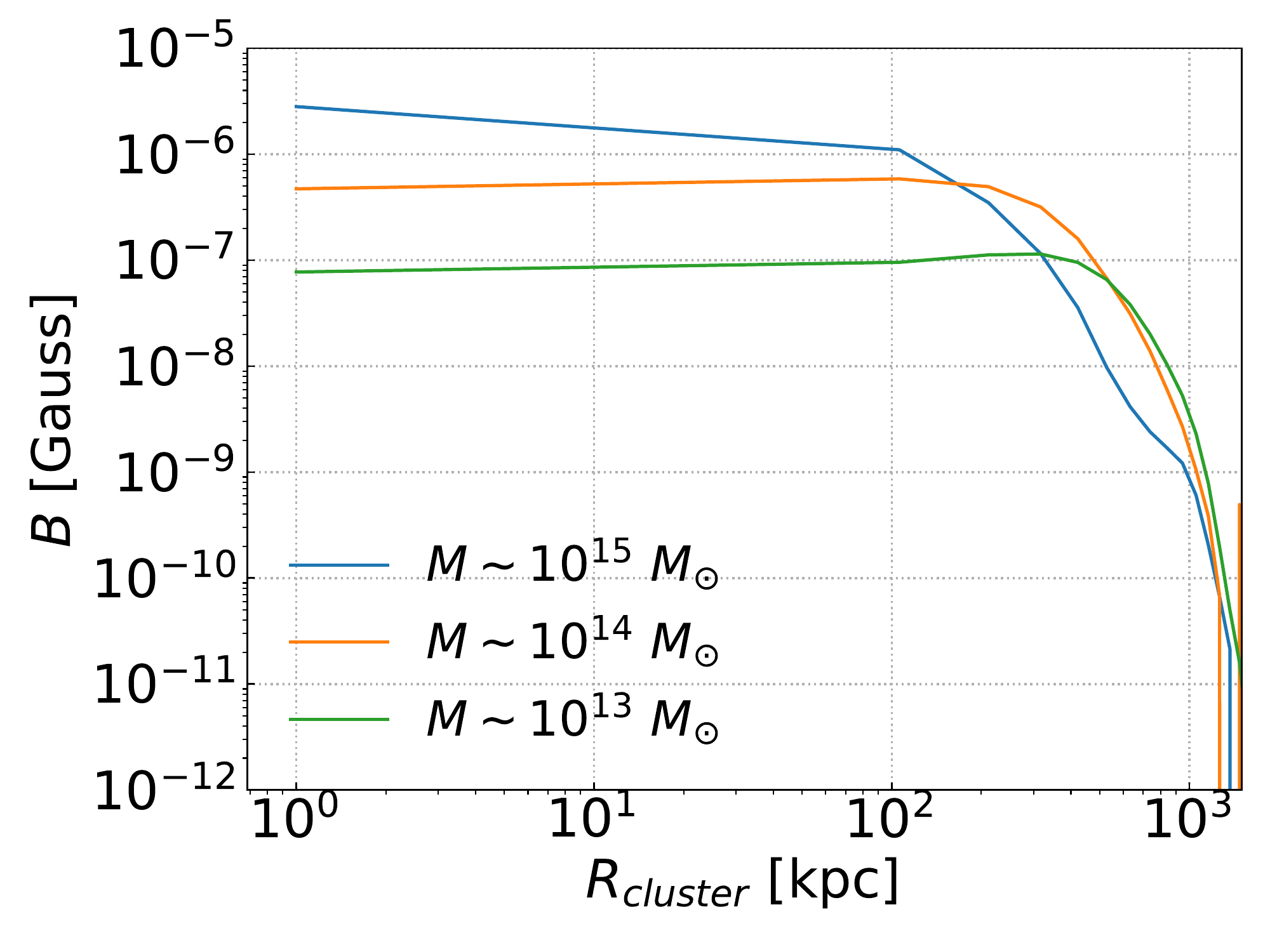}
\caption{Upper panel shows the whole volume-averaged value of the magnetic field as a function of the  cluster  mass. Lower panel compares  the volume-averaged magnetic field as a function of the radial distance for clusters of different masses.}
\label{fig:ClusterB}
\end{figure}

We also found that the magnetic field strength of a cluster depends on its mass: the heavier the cluster, the stronger the average magnetic field is, due to the larger extension of denser regions (see middle column of Fig.~\ref{fig:Cluster-contour} and Fig.~\ref{fig:ClusterB}). Inside all clusters, magnetic fields vary in the range $10^{-8} < B / \text{G} < 10^{-5}$ \citep[see also][]{dolag2005constrained, ferrari2008observations, xu2009turbulence,  brunetti2014cosmic, brunetti2017relativistic, brunetti2020second}. 

\begin{figure}
\centering
\includegraphics[width=\sizeFigM]{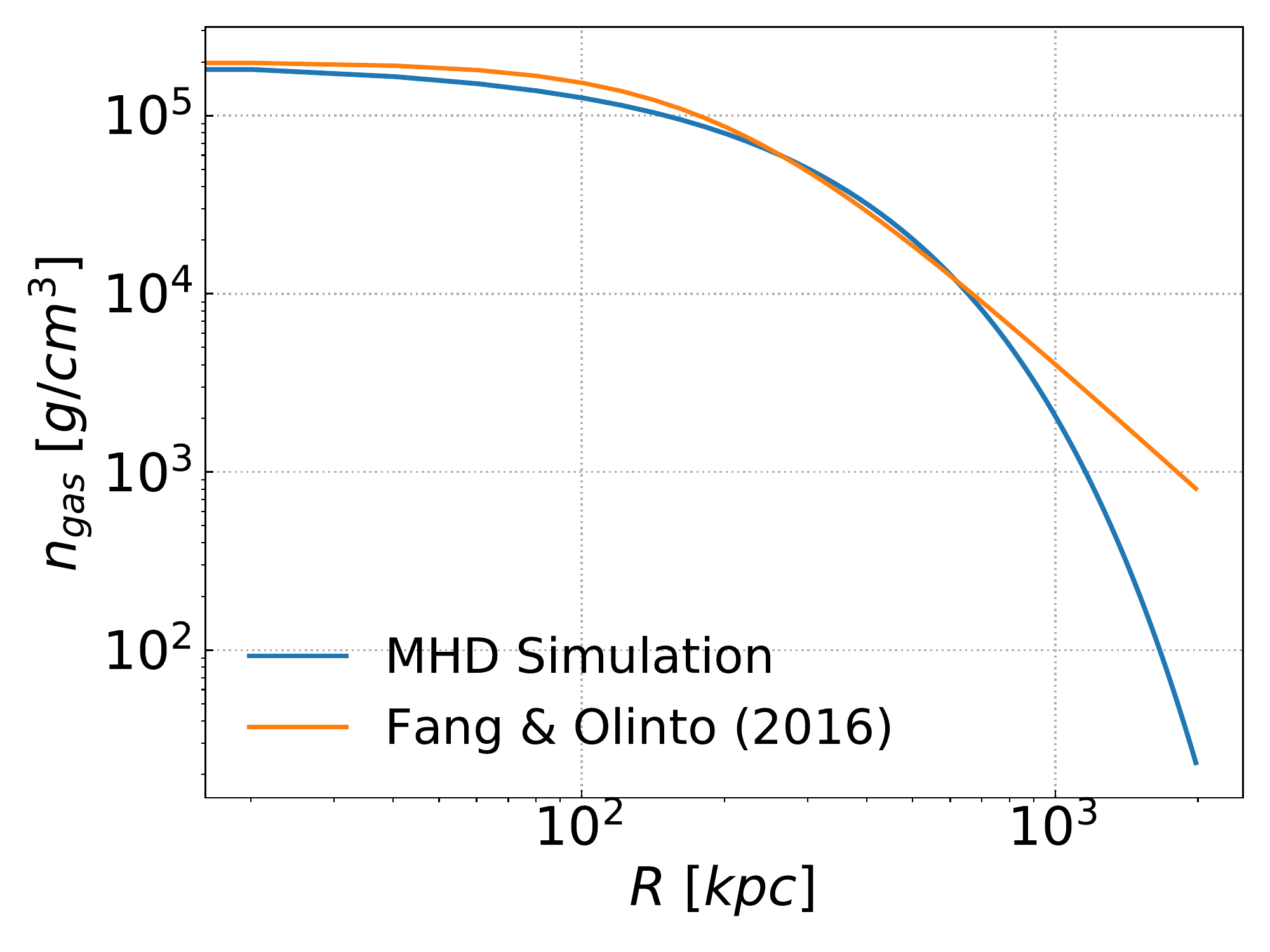}
\includegraphics[width=\sizeFigM]{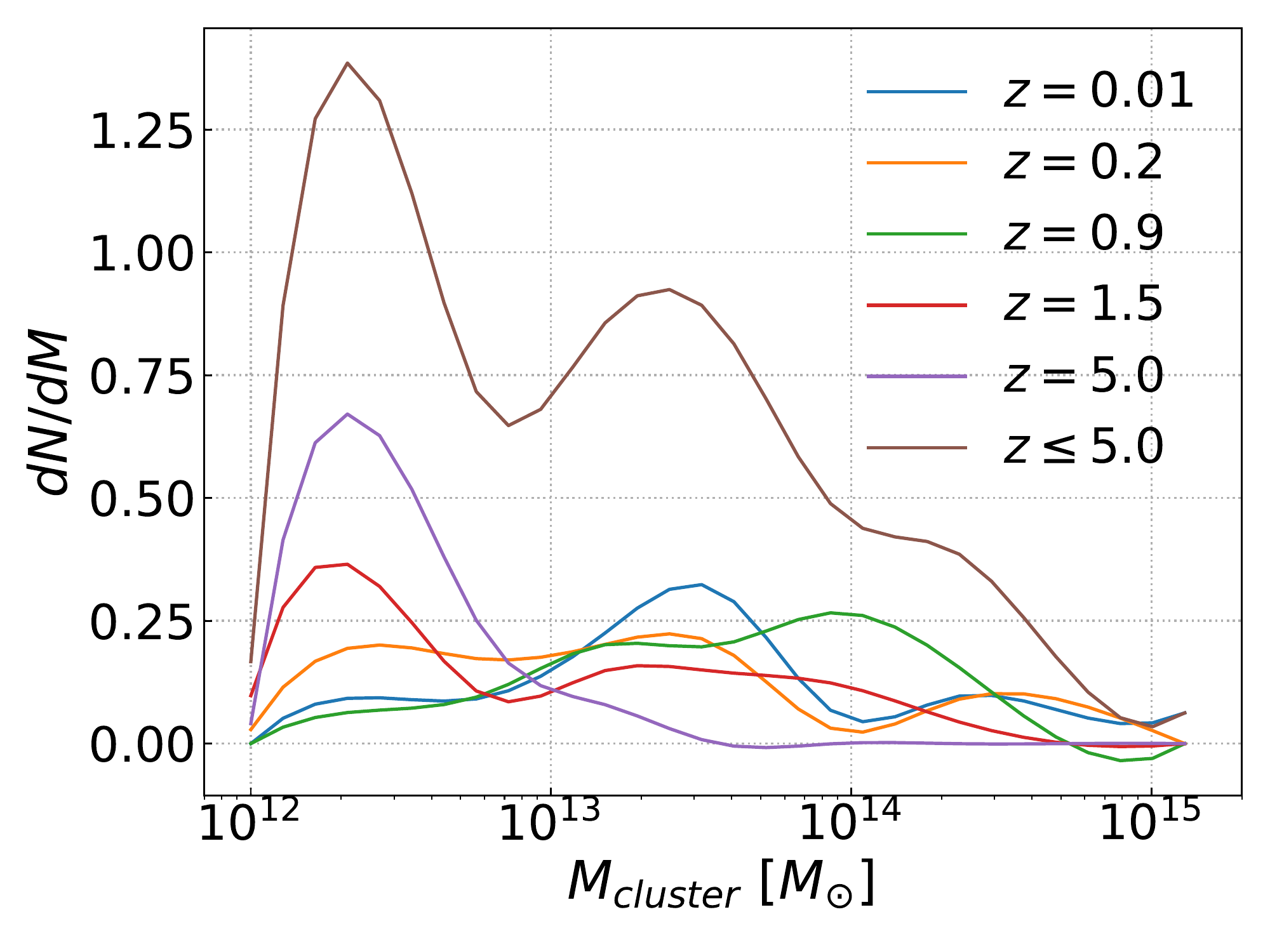}
\caption{Comparison of the density profile of a cluster of mass $10^{15} \; M_{\odot}$, from our simulation with the model used by \citet{fang2016high}, given in the upper panel. The lower panel shows the number of clusters per mass interval in our background simulation for different redshifts. }
\label{fig:ngas-massC-virgo}
\end{figure}

In the upper panel of Fig.~\ref{fig:ngas-massC-virgo}, we compare the radial density profile of our simulated  cluster of mass $10^{15} \; M_{\odot}$ with the model used by \citet{fang2016high}. We see that both profiles look similar up to $\sim 10^3 \; \text{kpc}$. Above this scale, the density distribution of our simulated clusters decays much faster than the assumed distribution in \citet{fang2016high}.

To estimate the total flux of CRs and neutrinos, we need to evaluate the total number of clusters in our background simulations as a function of their mass, at different redshifts.
From the entire simulated volume,  ($240 \; \text{Mpc})^3$,  we selected $20$ sub-samples of $(20 \; \text{Mpc})^3$ from different regions, as  representative of the whole background.
We then  calculated the average number of clusters per mass interval in each of these sub-samples ($dN_{\text{clusters, avg}} / dM$), between $10^{12} \; M_{\odot}$ and $10^{16} \; M_{\odot}$. 
To obtain the total number of clusters per mas interval we multiplied this quantity 
by the number of intervals $N = (240 \; \text{Mpc})^3 / (20 \; \text{Mpc})^3$ in which the whole volume was divided. So, the total number of clusters per mass interval was calculated as $(dN_{\text{clusters,\; avg}}/dM) \times N$.
Since we have seven redshifts in our cosmological background simulations, $z =  0.01, \; 0.05, \; 0.2, \; 0.5, \; 0.9, \; 1.5, \; 5.0$, we then have repeated the calculation above for each snapshot to obtain the number of clusters per mass interval at different redshifts. This is shown in the lower panel of Fig.~\ref{fig:ngas-massC-virgo} for different redshifts.

To calculate the photon field of the ICM, we assume that the clusters are filled with photons from Bremsstrahlung radiation of the hot, rarefied ICM gas (see Figs.~\ref{fig:filaments-Tmp}  to \ref{fig:Cluster_angular}). For typical temperatures and densities, we can further assume an optically thin gas. Taking a  photon density ($n_\text{ph}$) distribution with approximately spherical symmetric within the cluster, we have the following relations for an optically thin gas \citep{rybicki2008radiative}:

\begin{equation}\label{eq:phDen}
\frac{dn_\text{ph}}{d\epsilon} = \frac{4\pi I_{\nu}}{ch\epsilon}, \;\;\;\;\; I_{\nu} = R_\text{shell}~ J_{\nu}^\text{ff}.
\end{equation}
where $I_{\nu}$ is the specific intensity of the emission, $c$ is the speed of light, $h$ is the Planck constant, $\epsilon$ is the photon energy, $R_\text{shell}$ is the radius of concentric spherical shells, and $J_{\nu}^\text{ff}$ is related with the Bremsstrahlung emission coefficient: % \citep{rybicki2008radiative}:
\begin{equation} \label{eq:Emmisvity1}
4\pi J_{\nu}^\text{ff} = \epsilon_{\nu}^\text{ff} (\nu, n, T) =  6.8\times 10^{-38} Z^2 n_e n_i T^{-1/2}  e^{-h\nu/k_{B} T},
\end{equation}
which is given in units of  $\text{erg} \; \text{cm}^{-3} \; \text{s}^{-1} \; \text{Hz}^{-1}$.

\begin{figure}
\centering
\includegraphics[width=\sizeFigM]{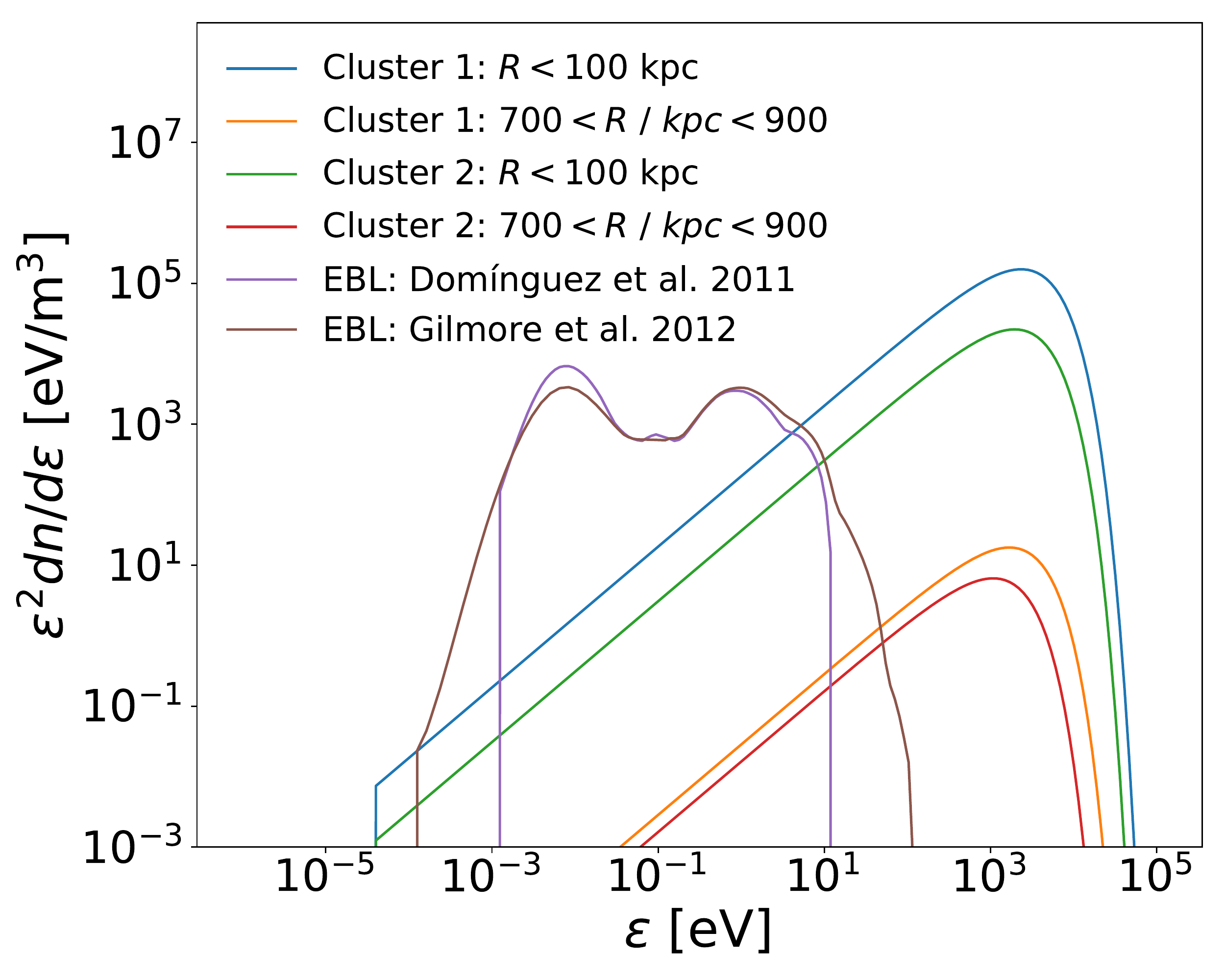}
\caption{Comparison of EBL with the Bremsstrahlung radiation of the ICM as a function of the photon energy. The Bremsstrahlung is calculated for two clusters at different radial distance intervals. Cluster$\; 1$ has mass $10^{15} \; M_{\odot}$, and 
Cluster$\;2$ , $10^{14} \; M_{\odot}$. }
\label{fig:EBL-Brem}
\end{figure}

In Fig.~\ref{fig:EBL-Brem} we compare the radiation fields for two EBL models \citep{gilmore2012semi, dominguez2011extragalactic} with the Bremsstrahlung photon fields of two clusters of masses $\sim 10^{15} \; M_{\odot}$ (cluster$\; 1$) and $\sim 10^{14} \; M_{\odot}$ (cluster$\;2$). For both clusters, we calculated the internal photon field at the center ($R < 100 \; \text{kpc}$) and for the ($700 < R \; / \; \text{kpc} < 900$). It can be seen that the Bremsstrahlung photon field is dominant at X-rays, but only near the center of the clusters, while the EBL dominates at infrared and optical wavelengths mainly. 

The interaction rates  of CRs with the Bremsstrahlung photon fields in each shell were also calculated (see Fig.~\ref{fig:MFP_Neu-Traj} and  appendix \ref{Appsec:MFP}) and implemented in CRPropa. We note that though the assumption of spherical symmetry  for evaluating the Bremsstrahlung radiation and its interaction rate with CRs seems to be in contradiction with the results of Fig.~\ref{fig:Cluster_angular}, our computation of these quantities in  CRPropa have revealed no significant contribution of the Bremsstrhalung photons to neutrinos production. Indeed, the upper panel of Fig.~\ref{fig:MFP_Neu-Traj}  indicates that the $\lambda$ for these interactions is larger than the Hubble horizon. Thus deviations from spherical  symmetry for this  photon field will not be relevant in this study.

We have also implemented the proton-proton (pp) interactions using the spatial dependent density field extracted directly from the background cosmological simulations, using the same procedure described by \citet{rodriguezramirez2019very}. We further notice that, for the computation of the CR fluxes,   the magnetic field distribution has been also extracted directly from the background  simulations, without considering any kind of space symmetry.

\subsection{Mean free paths for different CR interactions}\label{MFP}

CRPropa~3 employs a Monte Carlo method for particle propagation and previously loaded tables of the interaction rates in order to calculate the interaction of CRs with photons along their trajectories. We implemented the spatially-dependent interaction rates into the code, based on the gas and photon density distributions for the clusters of different masses. The mean free paths ($\lambda$) for the different interactions of CRs are described in appendix~\ref{Appsec:MFP}.

\begin{figure}
\centering
\includegraphics[width=\sizeFigM]{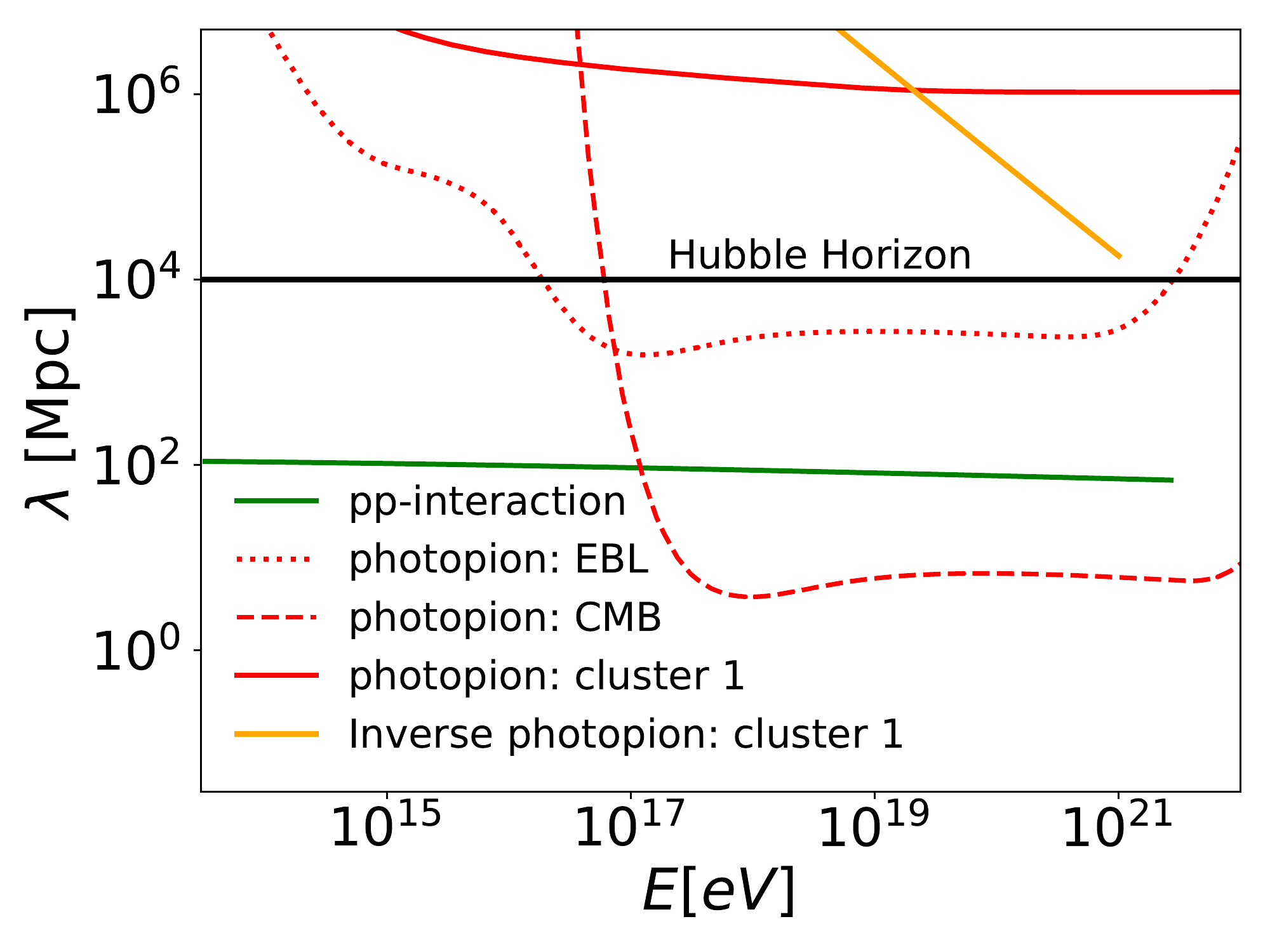}
\includegraphics[width=\sizeFigM]{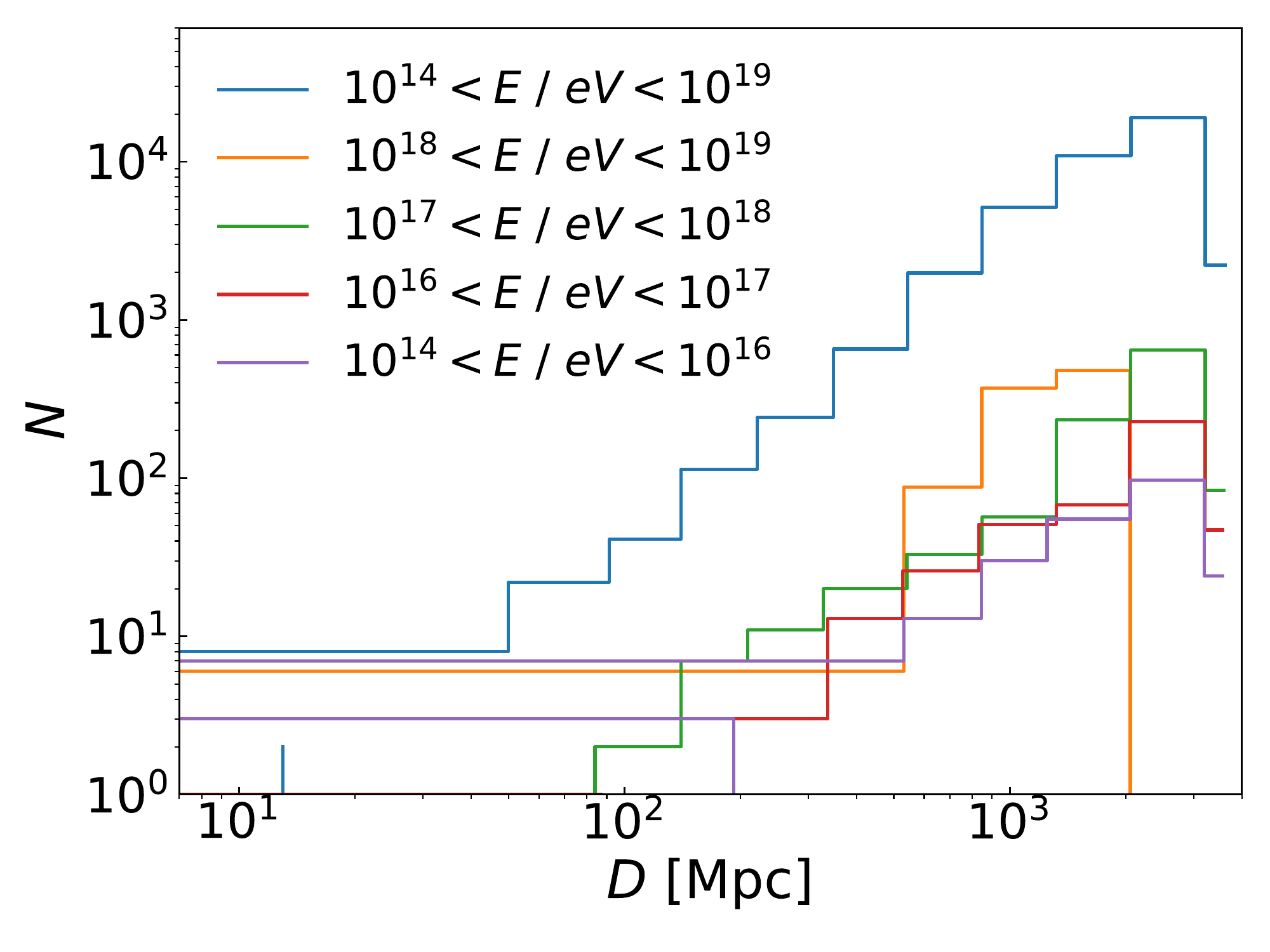}
\caption{The upper panel shows the mean-free path $\lambda$ for CR interactions which produce neutrinos. It is shown  $\lambda$ for  photopion production in the bremsstrahung photon field (red solid line), CMB (red dashed line) and EBL (red dotted line). Also shown is $\lambda$ for pp-interactions (green) calculated within a sphere of radius $r = 100$ kpc around the center of a massive cluster (with mass $10^{15} \; M_{\odot}$ and shown in  Fig.~\ref{fig:Cluster-contour}). The $\lambda$ for the
interaction of high-energy gamma rays with the local gas of the ICM (yellow) is also depicted. The thick black line represents the Hubble horizon in the upper panel.
The lower panel shows the distribution of the total trajectory length of CRs inside the cluster as a function of their energy bins.
}
\label{fig:MFP_Neu-Traj}
\end{figure}

The values of $\lambda$ for all the interactions of CRs with the background photon fields and the gas, are plotted in the upper panel of Fig.~\ref{fig:MFP_Neu-Traj}. For photopion production, we compare  $\lambda$ due to interactions with  the  photon fields (i.e., the Bremsstrahlung radiation, red solid line) of  a  cluster of mass $10^{15} \; M_{\odot}$   with the EBL  (red dotted line) and the CMB (red dashed line). For the Bremsstrahlung radiation,  we considered only the photons within a sphere of radius $100 \; \text{kpc}$ around the center of the cluster (i.e., the  densest region, which is shown in Figs.~\ref{fig:Cluster-contour} \&~\ref{fig:Cluster_angular}).
High-energy CR interactions with CMB photons is a well-understood process that limits the distance from which CRs can reach Earth leading to the GZK cutoff. The upper panel of Fig.~\ref{fig:MFP_Neu-Traj} shows that $\lambda$ for this interaction is much smaller than that for the EBL and Bremsstrahlung. 
So, CR interactions with CMB photons dominate at energies $E \gtrsim 10^{17} \; \text{eV}$. We also see  that  $\lambda$ for Bremsstrahlung is greater than the size of the universe ($\sim 10^{6}$ Mpc), and for  EBL, it is $\sim 10^{3} \; \text{Mpc}$. The $\lambda$ for  pp-interactions (green line)  is much less than the Hubble horizon. Therefore, this kind of interaction is  more likely to occur than photopion production specially at energies $< 10^{17}$ eV. 
Upper panel of Fig.~\ref{fig:MFP_Neu-Traj} also shows  that we can neglect the CR interactions with the local Bremsstrahlung photon field,  as well as the interaction of high-energy gamma rays with the local gas of the ICM (yellow) in photopion production.

The lower panel of Fig.~\ref{fig:MFP_Neu-Traj} shows the distribution of the trajectory lengths (total distance travelled by a CR inside the cluster up to the observation time), for different energy bins of CRs. There is a substantial  number of events with trajectory length greater than $D \gtrsim 10^{3}$ Mpc for each energy bin. Thus, the trajectory lengths of CRs are comparable to the mean free paths of pp-interactions and photopion production in the CMB and EBL case, so that these interactions can produce secondary particles including gamma rays and neutrinos.

\subsection{CR Flux Calculation }\label{sec:CRs-Spec}

To study the propagation of CRs in the diffuse ICM, we used the transport equation as implemented in CRPropa 3 by \citep[][see also equation \ref{b1}]{merten2017crpropa}. There are three possible scenarios in CRPropa3 for each particle until its detection: the particle reaches the detector within a Hubble time; the energy of the particle becomes smaller than a given threshold; or the trajectory length of a  CR exceeds the maximum propagation distance allowed.

We inject CRs isotropically  with a power-law energy distribution with spectral index $\alpha$ and exponential cut-off energy $E_{\text{max}}$ which follows the relation
$dN_{\text{CR}, E}/dE \propto E_i^{-\alpha} \exp(-E_i/E_{\text{max}})$ 
(see Appendix \ref{Appsec:specIndex}).
We  take different values for $\alpha \simeq 1.5 - 2.7$, and for  $E_{\text{max}} = 5\times10^{15} - 10^{18}$ eV \citep[e.g.][for review]{brunetti2014cosmic, fang2016high, brunetti2017relativistic, hussain2019propagation}. 

As stressed, the lower and upper limits of the mass of the  galaxy clusters are taken to be $10^{12} \ M_{\odot}$ and $10^{16} \; M_{\odot}$, respectively. This is because 
for $10^{14} \lesssim E / \text{eV} \lesssim 10^{19}$,
clusters with mass $M < 10^{12} \; M_{\odot}$ barely contribute to the total flux of neutrino, due to low gas density, 
while there are few clusters with $M \gtrsim 10^{15} \; M_{\odot}$ at high redshifts ($z > 1.5$)
\citep{komatsu2009five, ade2014planck}. The closest galaxy clusters  are located at $z\sim 0.01$, so we consider the  redshift range $0.01 \leq  z \leq 5.0$.  

The amount of power of the clusters that goes into CR production is left as a free parameter to be regulated by the observations \citep[e.g.][for reviews]{gonzalez2013galaxy, brunetti2014cosmic, fang2016high}.
We here assume 
that about $0.5 - 3 \; \%$ of the cluster luminosity is available for particle acceleration.  

%\textcolor{red}{

We did not consider the feedback from active galactic nuclei (AGN) or  star formation rate (SFR) in our background cosmological simulations \citep[as performed e.g. in][]{barai_etal2016, barai_dalpino2019}. AGN are believed to be the most promising CR accelerators inside clusters of galaxies and star-forming galaxies  contain many supernova remnants that can also accelerate CRs up to very-high energies ($E \gtrsim 100 \; \text{PeV}$) \citep{he2013diffuse}.
AGN are more powerful and more numerous at higher redshifts \citep{hasinger2005luminosity, khiali_dalpino2016, d2020dust}, and their luminosity density evolves more strongly for $z \gtrsim 1$.
Also, supernovae are more common at high redshifts \citep{he2013diffuse, moriya2019first}. 
Therefore, it is reasonable to expect that, if high energy cosmic ray (HECR) sources have a cosmological evolution similar to AGN or following the star-formation rate (SFR), then the flux of neutrinos may be higher at high redshifts due to the larger CR output from these objects.

% \textcolor{orange}{[Saqib: Notation is still inconsistent. You problably didn't pay attention to my last email.]}
% \textcolor{red}{
% \textcolor{magenta}{\textbf{Saqib, I do not understand the sentence below in the [ ]}}
% \textbf{The evolution} of different sources \citep[][]{hasinger2005luminosity, gelmini2012gamma, batista2019cosmogenic} can be determined by the X-rays luminosity evolution $L_{\text{X}}(z)$ of sources at redshift $z$ as function of the present
% luminosity $L_{\text{X},\; 0}$, as
% $L_{\text{X}}(z) \propto L_{\text{X}, \; 0}\; (1+z)^n \propto L_{\text{X}, \; 0} \; \psi(z)$.

%\textcolor{red}{
For the evolution of  AGN sources  \citep{hopkins2006normalization, heinze2016cosmogenic} and SFR \citep{yuksel2008revealing, wang2011constraining, gelmini2012gamma}  we consider the following parametrization:
\begin{equation}\label{eq:SFRz}
\psi_{\text{SFR}}(z) = \frac{1}{B} 
\begin{cases} 
(1+z)^{3.4} \;\; \text{if} \; z < 1 \\
(1+z)^{-0.3} \;\; \text{if} \; 1 < z < 4 \\
(1+z)^{-3.5} \;\; \text{if} \; z > 4 \\
\end{cases}
\end{equation}
\begin{equation}\label{eq:AGNz}
\psi_{\text{AGN}}(z) = \frac{(1+z)^m}{A} 
\begin{cases}
(1+z)^{3.44} \;\; \text{if} \; z < 0.97 \\
10^{1.09}(1+z)^{-0.26} \;\; \text{if} \; 0.97 < z < 4.48 \\
10^{6.66}(1+z)^{-7.8} \;\; \text{if} \; z > 4.48 \\
\end{cases}
\end{equation}
where $A = 360.6$ and $B = 6.66$ are normalization constants in equations~(\ref{eq:AGNz}) and~(\ref{eq:SFRz}), respectively.
For AGN evolution $\psi_{AGN}(z) \propto (1+z)^5$, for low redshift $z < 1$ \citep{gelmini2012gamma, batista2019cosmogenic} and also according to \citep{gelmini2012gamma, heinze2016cosmogenic}, in equation (\ref{eq:AGNz}), $m > 1.5$ for AGN, so we consider $m=1.7$.
Typically, the luminosity of AGNs ranges from $10^{42}$ to $10^{47} \; \text{erg/s}$ and their evolution depends on their luminosities.
The AGNs with luminosities $\sim 10^{44} - 10^{46} \; \text{erg/s}$ are more important as they are more numerous and  believed to be able to accelerate particles to ultra-high energies \citep[e.g.][]{waxman2004extra, khiali_dalpino2016}.  AGNs with luminosities greater than $10^{46} \; \text{erg/s}$ are less numerous \citep{hasinger2005luminosity}  and their evolution function ($\psi_{\text{AGN}}(z)$) is different from equation (\ref{eq:AGNz}).
For no source evolution, $\psi(z) = 1$.

The total flux of CRs is  estimated from the entire population of clusters. The number of clusters per mass interval $dN/dM$ at redshift $z$ is given in the lower panel of Fig.~\ref{fig:ngas-massC-virgo}, which was obtained from our cosmological simulations. It is related to the flux through:
\begin{equation}\label{eq:flux_calc}
E^{2} \Phi(E) = \int\limits_{z_\text{min}}^{z_\text{max}} dz \int\limits_{M_\text{min}}^{M_\text{max}} dM \dfrac{dN}{dM}~ E^{2}  \dfrac{d\dot{N}( E/(1+z), M , z)}{dE}
\left( \dfrac{\psi_{\text{ev}}(z)}{4\pi d_L^2 (z)}\right) 
\end{equation}
where $\psi_{\text{ev}}(z)$ stands for, $\psi_{\text{SFR}}(z)$ and $\psi_{\text{AGN}}(z)$,   $\dot{N}$ is the  number of CRs 
per time interval $dt$ 
with energies between $E$ and $E + dE$ that reaches the observer. The quantity $E^2 \; d\dot{N}/dE$ in equation (\ref{eq:flux_calc}) is  the power of CRs calculated 
from our propagation simulation and  is several orders of magnitude smaller than the luminosity of observed clusters \citep[e.g.,][]{brunetti2014cosmic}.

In order to convert the code units of the CR simulation to  physical units, we have used a  normalization factor (Norm).
To calculate Norm, we first evaluate  the X-ray luminosity of the cluster using the empirical relation $L_{\text{X}} \propto f_g^2 \; M_{\text{vir}}$ \citep[][]{schneider2014extragalactic}, where $f_g = M_g/M_{\text{vir}}$ denotes the gas mass ($M_g$) fraction  with respect to the total mass of the cluster within the Virial radius ($M_{\text{vir}}$) and then, since we are assuming that $(0.5 - 3) \; \%$ of this luminosity goes into CRs, this implies that    $\text{Norm} \sim (0.5 - 3) \; \%~L_{\text{X}}/L_{\text{CRsim}}$ and $L_{\text{CRsim}}$ is the luminosity of the simulated CRs. Therefore, the CR power that reaches  the observer (at the Earth)  is $\sim E^2~d\dot{N}/dE \times \text{Norm}$. In equation (\ref{eq:flux_calc}) $d_L$ is the luminosity distance, given by: 
\begin{equation}
    d_L = (1 + z) \dfrac{c}{H_0} \int\limits_{0}^{z} \frac{dz^\prime}{E(z^{\prime})},
\end{equation}
with
\begin{equation}
    E(z) = \sqrt{ \Omega_m(1+z)^3  + \Omega_\Lambda}  = \frac{H(z)}{H_0},
\end{equation}
where the Hubble constant, as well as the matter ($\Omega_m$) and dark-energy ($\Omega_\Lambda$) densities are defined in section \ref{sec:MHD-Sim}, assuming a flat $\Lambda$CDM universe.

\begin{figure}
\centering
\includegraphics[width=\sizeFigM]{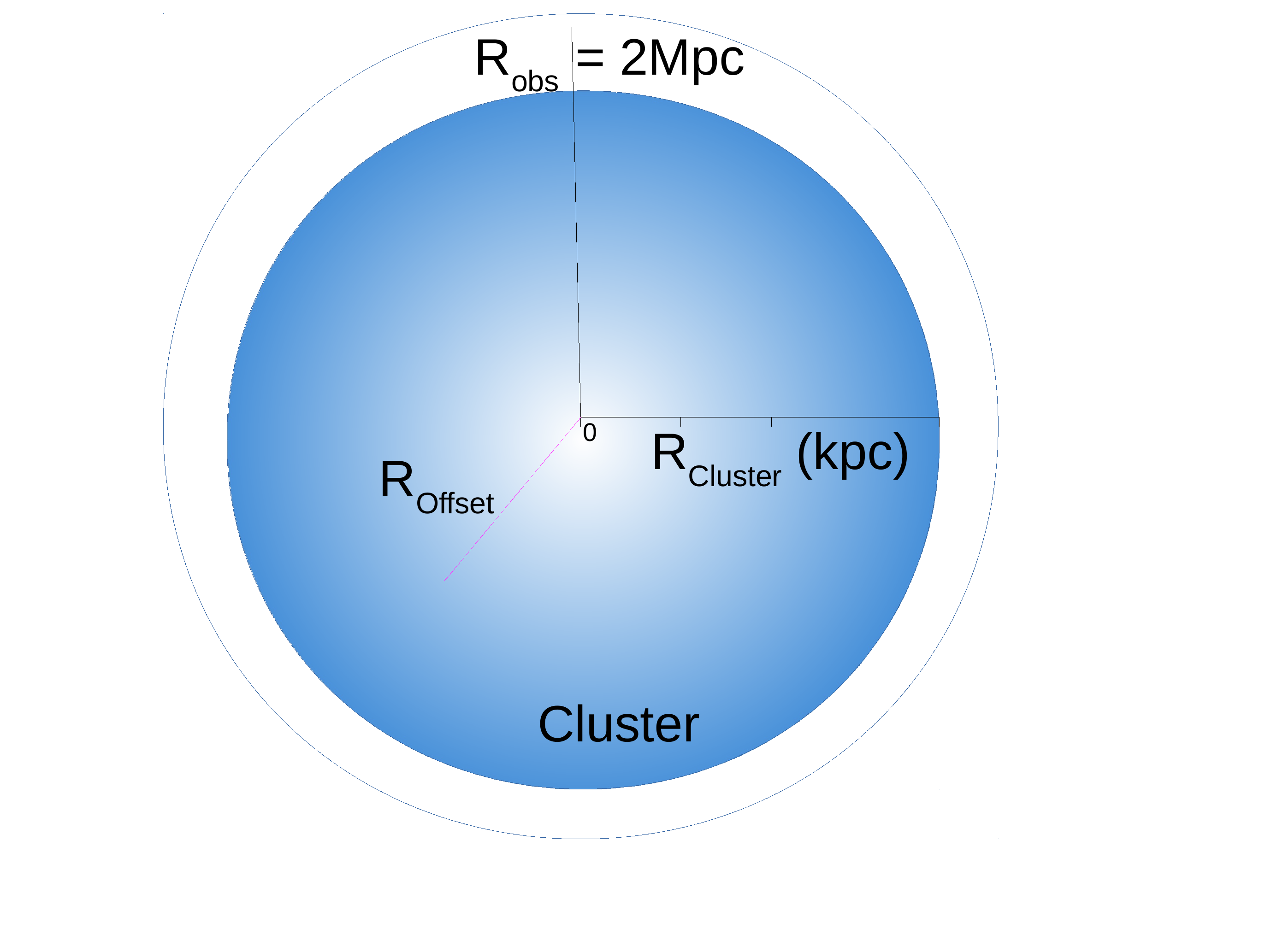}
\caption{Scheme of the CR simulation geometry.
They are injected at three different positions inside each cluster represented by $R_{\text{Offset}}$, and $R_{\text{Obs}}$ is the radius of the observer.}
\label{fig:schematic-diagram}
\end{figure}

We selected different injection points inside the clusters of different masses in order to study the spectral dependence  with the position,  which may  correspond to different scenarios of acceleration of CRs.  For  instance, the larger concentration of galaxies near the center must favor more efficient acceleration, but compressed regions by shocks in the outskirts may also accelerate CRs.
The schematic diagram of the simulation of CRs propagation is shown in Fig.~\ref{fig:schematic-diagram}.
CRs are injected at three different positions within each selected cluster denoted by $R_{\text{Offset}}$.
The spectra of CRs have been collected by an observer in a sphere of $2~\text{Mpc}$ radius ($R_{\text{Obs}}$), centred at the cluster, with a redshift window ($-0.1 \leq z \leq 0.1$) for all the  injection points of CRs.
All-flavour neutrino fluxes are also computed at the same observer (see Section-\ref{sec:Neuspec} below).

\begin{figure}
\centering
\includegraphics[width=\columnwidth]{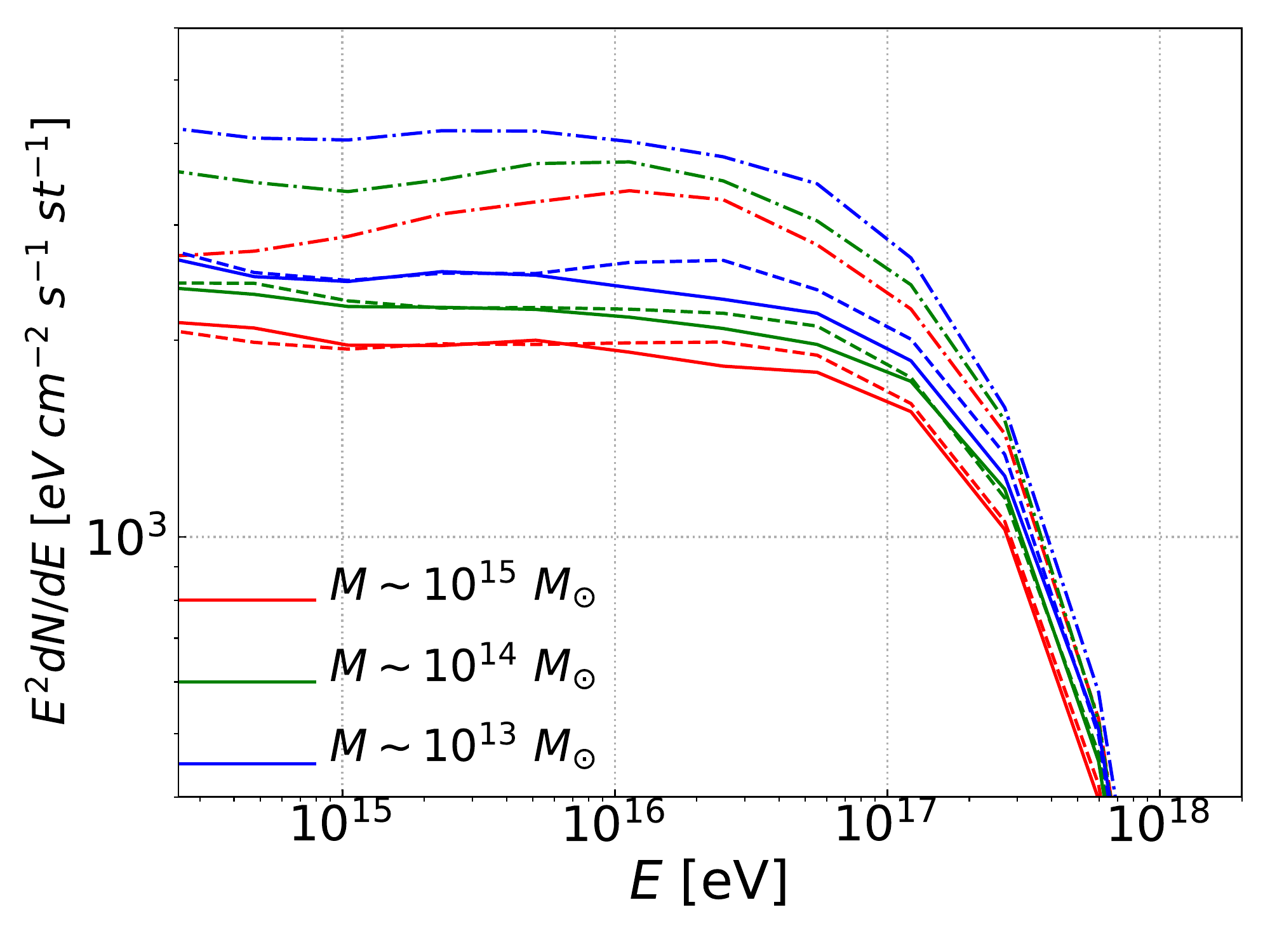}
\caption{This figure shows the CR flux of individual clusters of distinct masses,  $M\sim 10^{15}$ (red);  $10^{14}$ (green); and $M\sim 10^{13}~M_{\odot}$ (blue color).  This diagram shows  the flux of CRs, for  sources located at 
the center of the cluster (solid), at  $300$ kpc (dashed), and at $1$ Mpc (dash-dotted lines) away from the centre. The flux is computed at the edge of the clusters.  The spectral parameters are $\alpha=2$ and $E_{\text{max}} = 5\times 10^{17}$ eV, and it is assumed that  $2\%$ of the luminosity of the clusters is converted into CRs. }
\label{fig:CRs_Ms}
\end{figure}

\begin{figure}
\centering
\includegraphics[width=\columnwidth]{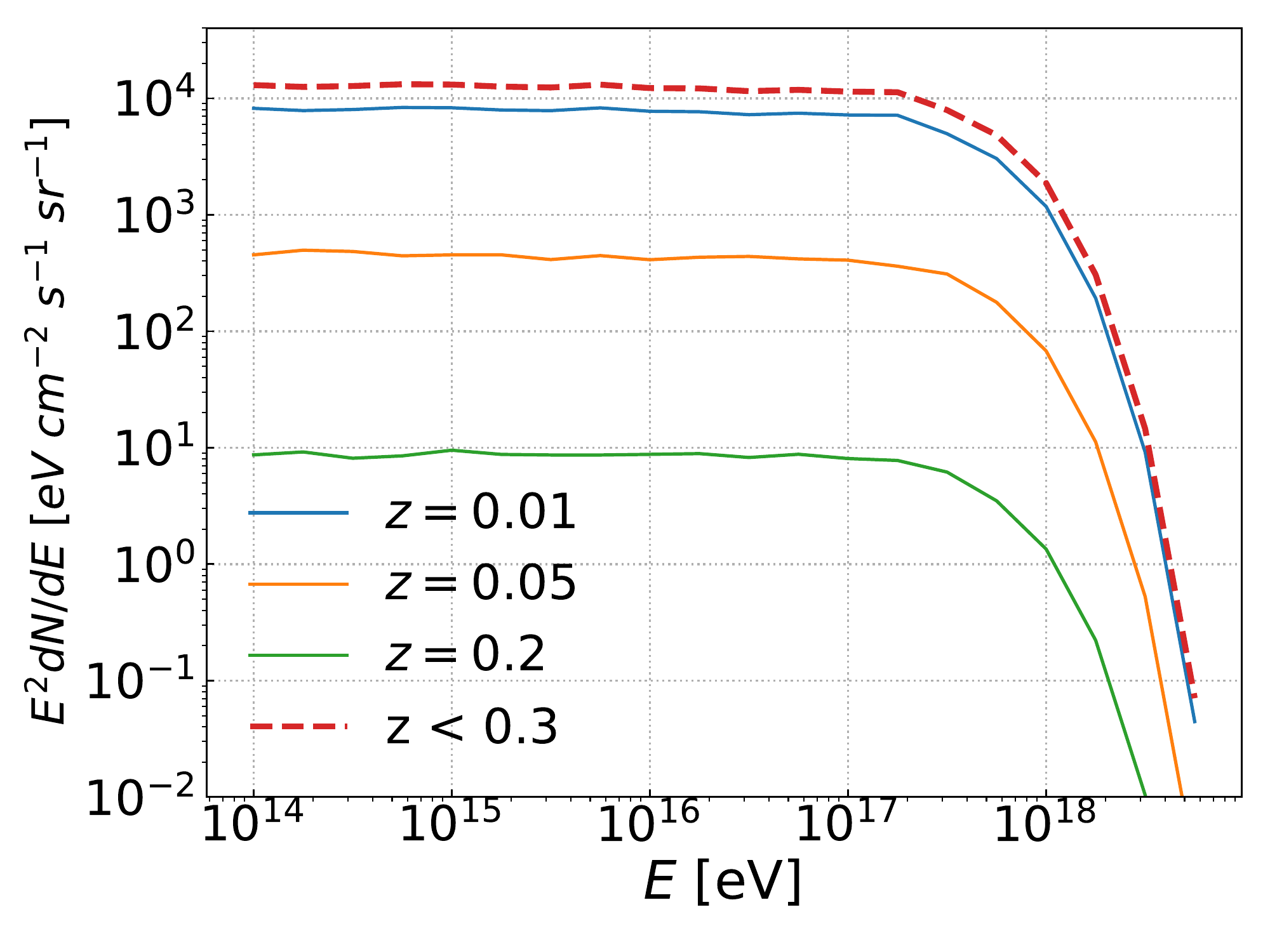}
\caption{ This figure shows  the total CR flux (at the Earth distance) from all the clusters distributed in different redshifts: $z =0.01$ (blue);  $z=0.05$ (orange); \ $z=0.2$ (green). The total CRs flux for the redshift range $0.01 \leq z \leq 0.3$ is given by the red dotted line. }
\label{fig:CRs_total}
\end{figure}

The spectrum of CRs obtained from our simulations is shown in Figs.~\ref{fig:CRs_Ms} \& \ref{fig:CRs_total}. 
Its dependence on the position where the CR source is located within the cluster
for $z=0.01$ 
is shown for three clusters of different masses in Fig.~\ref{fig:CRs_Ms}.
Particles injected at $1~\text{Mpc}$ distance away from the clusters center can leave them in short time, with  almost no interaction, as both the magnetic field and the gas number density are very low compared to the central regions.
 On the other hand, CRs injected at the center or at $300~\text{kpc}$ away from the cluster center can be easily deflected   by the magnetic field and trapped in  dense regions. This explains the higher CR flux for the injection point at $1~\text{Mpc}$ in Fig.~\ref{fig:CRs_Ms}. 
Also, because the confinement of CRs in the central regions of the clusters is comparable to a Hubble time, and because of the value of  $\lambda$ for the relevant interactions, the production of secondary particles including neutrinos and gamma rays in the clusters is substantial, as we will see in section \ref{sec:Neuspec}.

%%%%%%%%%%%%%%%%%%%%%

In Fig.~\ref{fig:CRs_total} we show the CR spectrum of all the clusters  at different redshifts integrated up to the Earth. 
Although the spectra in this diagram have been integrated up to the Earth,  we have not considered any interactions of the CRs with the background photon  and magnetic fields during their propagation from the edge of the clusters to the Earth. Though not quantitatively realistic,  it  provides important qualitative information. One  obvious result is that most of the  contribution in the CR flux comes from clusters at low redshifts. Moreover there is a significant suppression in the flux of CRs at $\gtrsim 10^{17}$~eV, which indicates the trapping of lower-energy CRs within the clusters~\citep{batista2018cosmic}.

%%%%%%%%%%%%%%%%%%%%%%%%%
%%%%%%%%%%%%%%%%

\subsection{Flux of Neutrinos }\label{sec:Neuspec}

To calculate the neutrino flux, the CRPropa 3 code integrates  a relation similar to  equation (\ref{eq:flux_calc}) for neutrino species, and the procedure is the same as described in Section \ref{sec:CRs-Spec} .

In general, neutrino production occurs mainly due to photopion production and pp-interactions.
In Fig.~\ref{fig:MFP_Neu-Traj}, where we show  $\lambda$ for different interactions, we see that  protons with energies
$E<10^{17}$ eV produce neutrinos principally due to pp-interactions, while  for $E>10^{17}$ eV, they produce neutrinos both, by pp-interactions and photopion process.
 We have also seen in Fig.~\ref{fig:MFP_Neu-Traj} (lower panel) that  the total trajectory length of CRs inside a cluster is comparable or larger than $\lambda$  for these interactions 
and thus, neutrino production is  inevitable.

\begin{figure}
\centering
\includegraphics[width=\columnwidth]{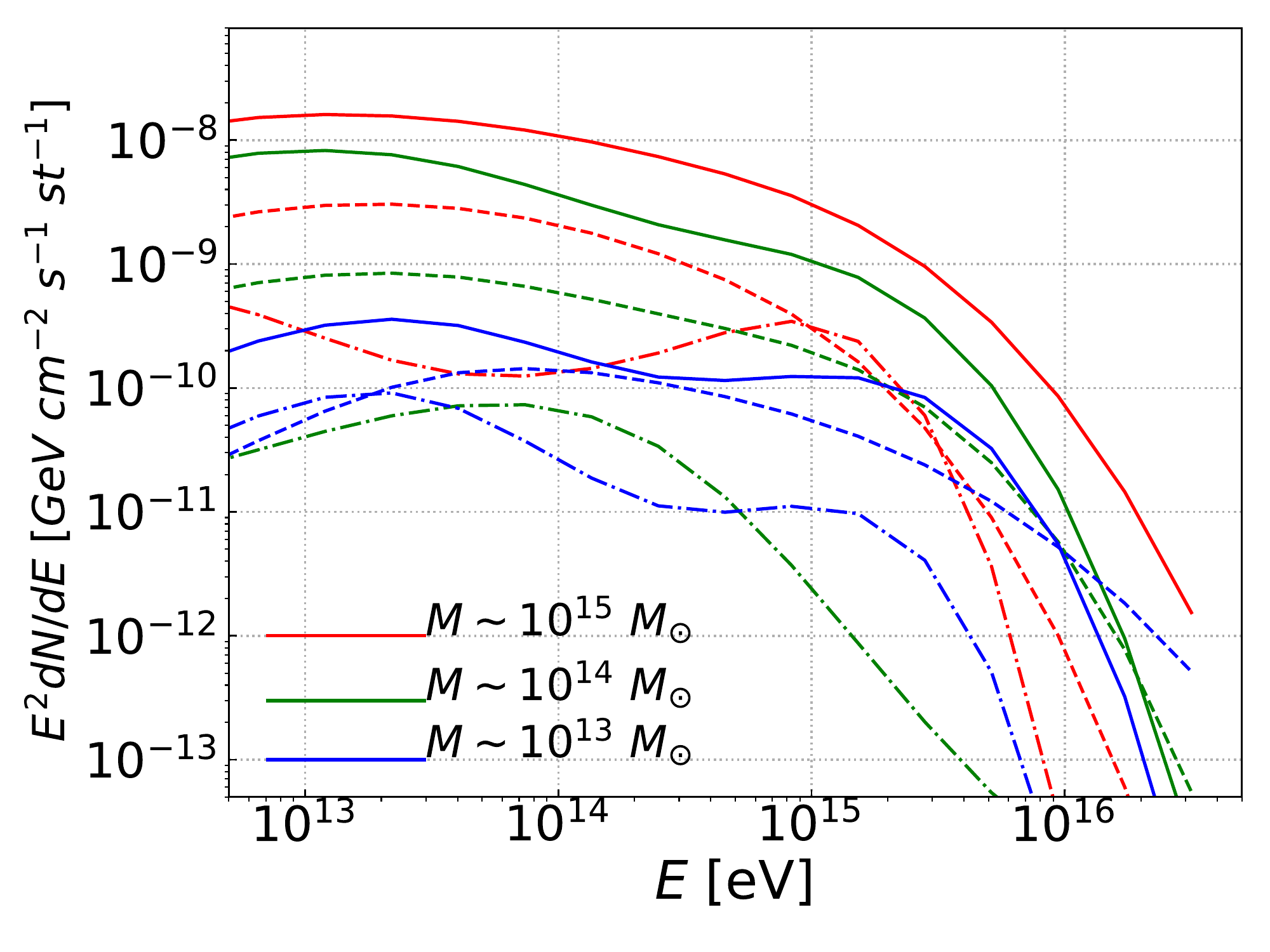}
\caption{This figure shows the neutrino flux of individual clusters of distinct masses:  $M\sim 10^{15}$ (red); $10^{14}$ (green) and $10^{13}~M_{\odot}$ (blue color). The CR  sources are located at the center of the cluster (solid lines), at  $300$ kpc (dashed lines), and at $1$ Mpc away from the center (dash-dotted lines). The flux is computed at the edge of clusters.  The CR injection follows $dN/dE \propto E^{-2}$, $E_{\text{max}} = 5\times 10^{16}$ eV, and it is assumed that  $2~\%$ of the luminosity of the clusters is converted to CRs. }
\label{fig:Neu-M}
\end{figure}

\begin{figure}
\centering
\includegraphics[width=\sizeFigM]{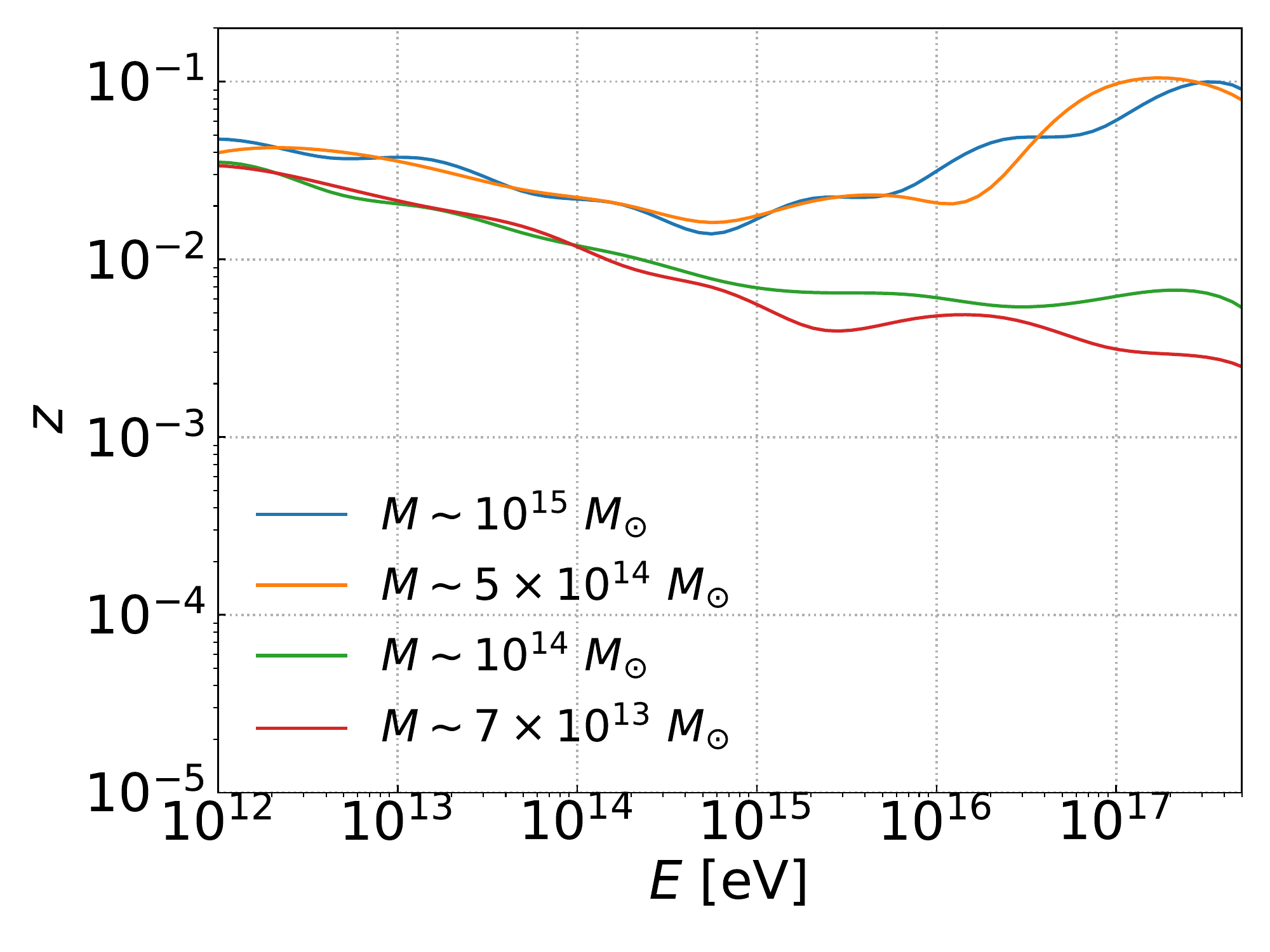}
\caption{Redshift distribution of the neutrinos as a function of their energy, as observed at $2$ Mpc away from the center of clusters with different masses.}
 \label{fig:Neu_zCluster} 
\end{figure}

In Fig.~ \ref{fig:Neu-M} we show the dependence of  the neutrino flux with the position of the corresponding CR source within clusters of different masses. As in the case of the CR flux, it can be seen that there is less neutrino production for the injection position at $1$~Mpc away from the center of the cluster. Furthermore, massive clusters produce more neutrinos than the light ones. In Fig.~\ref{fig:Neu_zCluster} we present the redshift distribution of neutrinos as a function of their energy, as observed at a distance of $2$~Mpc from the center of individual clusters with different masses.

%%%%%%%%%%%%%%%%%%%%%%%%%%%%%%%%%%%%%%%%%
%%%%%%%%%%%%%%%%%%%%%%%%%%%%%%%%%%%%555

\begin{figure*}
\centering
\includegraphics[width=\sizeFigL]{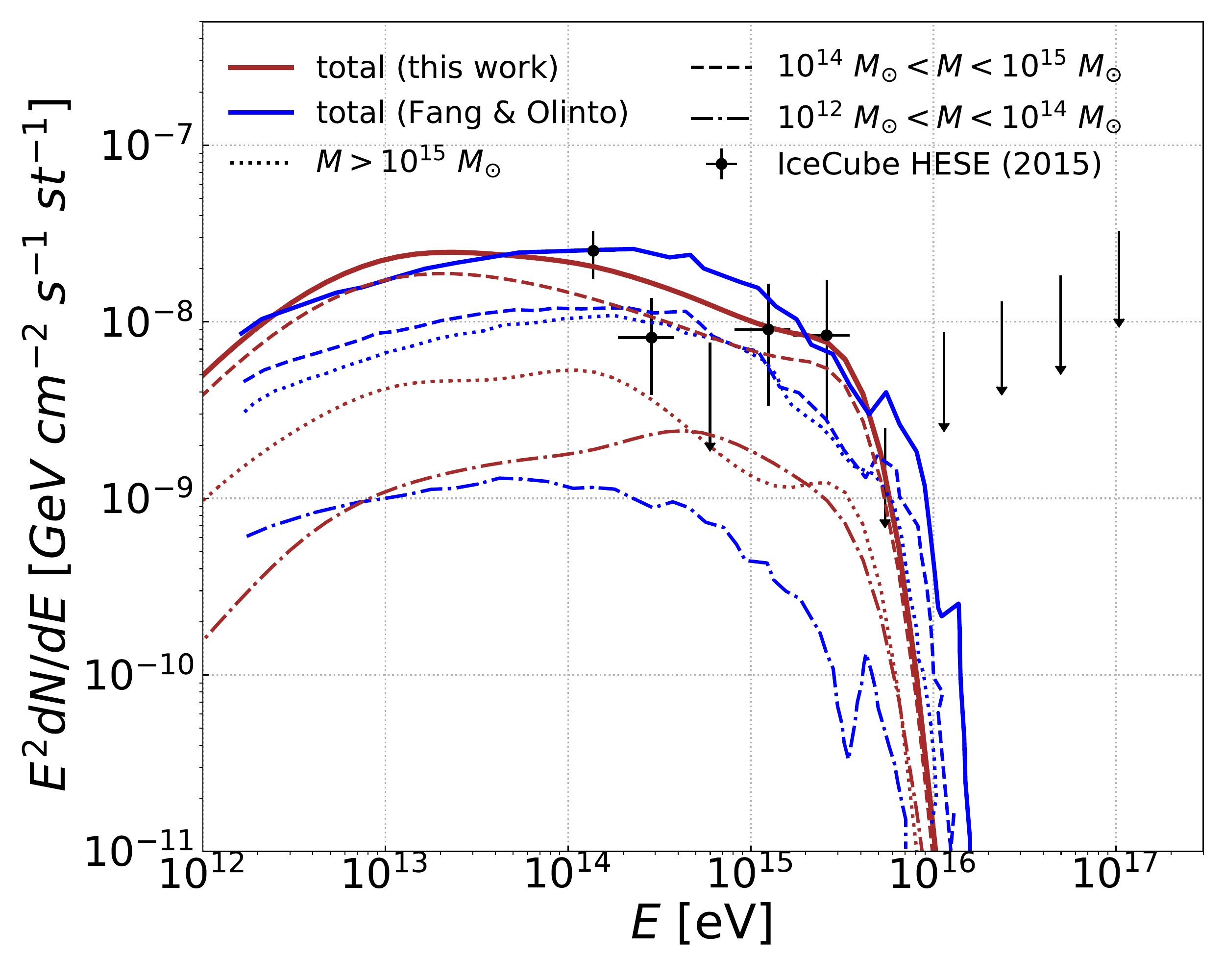}
\includegraphics[width=\sizeFigL]{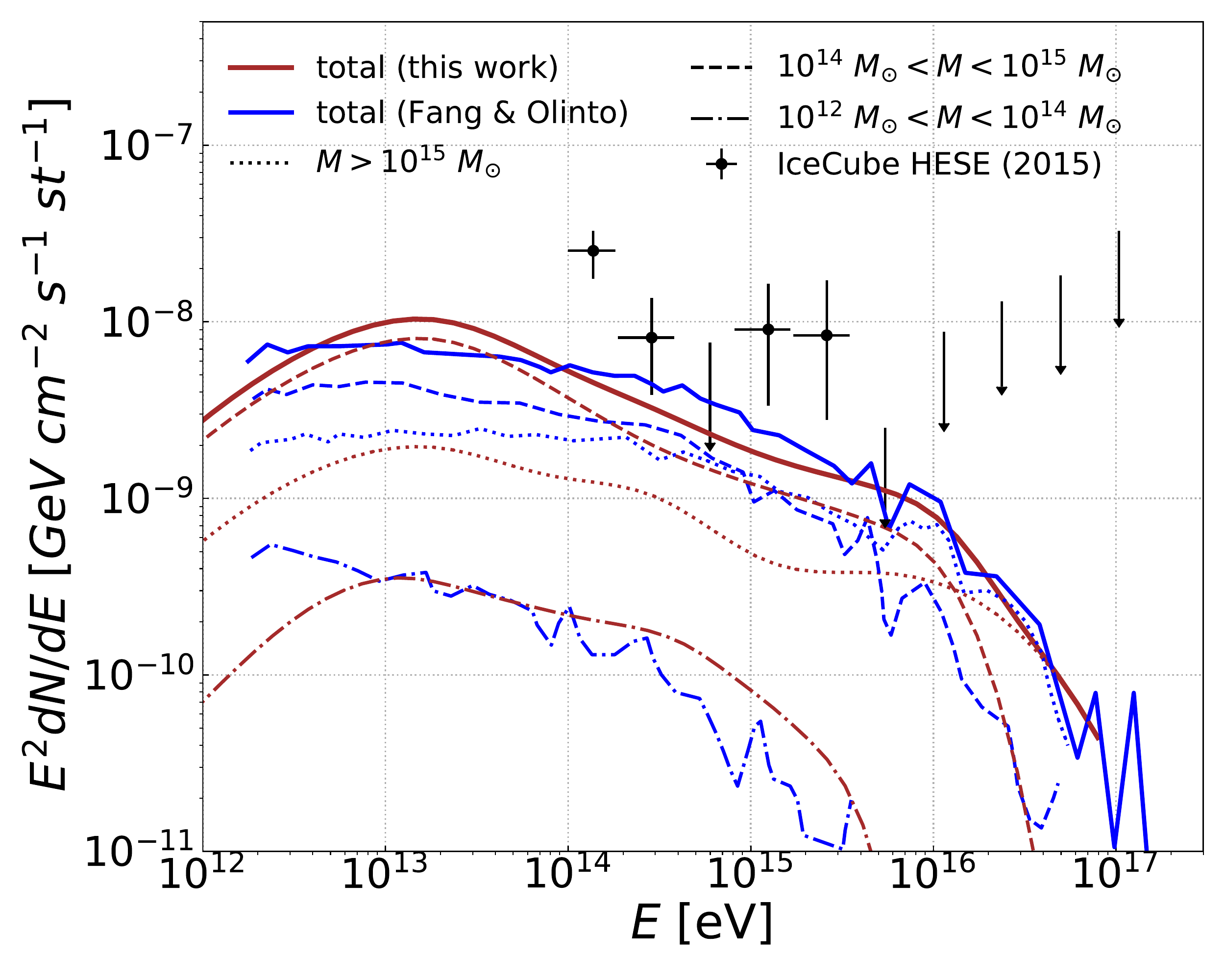}
\caption{Neutrino spectrum at Earth obtained using our simulations (brown lines), compared with the IceCube data (markers), and \citet{fang2016high} results (blue lines). The panels show the total flux integrated over all clusters and redshifts  between $0.01 \leq z \leq 5$ (solid thick lines). The left panel shows the neutrino spectra (thin blue and brown lines) for  cluster mass ranges of: $10^{12}~M_{\odot} < M < 10^{14}~M_{\odot}$ (dash-dotted), $10^{14}~M_{\odot} < M < 10^{15}~M_{\odot}$ (dashed), and $M > 10^{15}~M_{\odot}$ (dotted lines). The left panel corresponds to the case with $\alpha=1.5$ and $E_{\text{max}} = 5 \times 10^{16}$~eV, whereas in the right panel $\alpha = {-2}$ and $E_{\text{max}}$ follows equation (\ref{eq:emax}). 
These diagrams do not include the redshift evolution of the CR sources, $\psi_{ev}=1$ in  equation
equation (\ref{eq:flux_calc}). 
}
\label{fig:Neu_FangSimM}
\end{figure*}

\begin{figure}
\centering
\includegraphics[width=\sizeFigL]{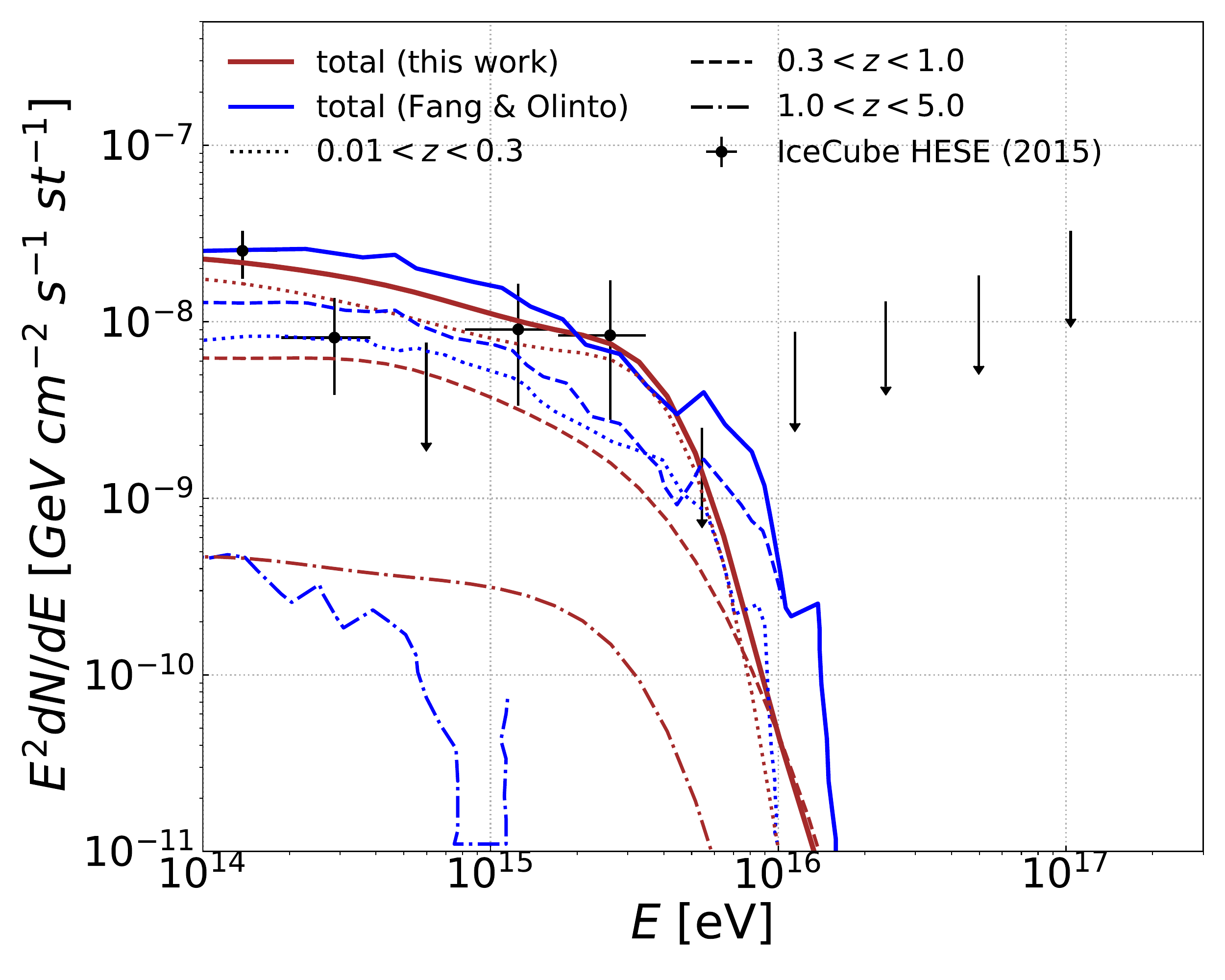}
\caption{This figure shows the neutrino spectrum for different redshift ranges: $z < 0.3$ (dotted lines), $0.3 < z < 1.0$ (dashed), and $1.0 < z < 5.0$ (dash-dotted lines). The solid blue and brown lines correspond to the total spectrum in \citet{fang2016high}, and in this work, respectively. The CR injection in this figure follows $dN/dE \propto E^{-1.5}$, and $E_\text{max} = 5 \times 10^{16} \; \text{eV}$. 
This figure does not include the redshift evolution of the CR sources, $\psi_{ev}=1$ in  equation
equation (\ref{eq:flux_calc}). 
}
\label{fig:Neu_FangSimz}
\end{figure}

In Fig.~\ref{fig:Neu_FangSimM} \&~ \ref{fig:Neu_FangSimz}, we present the total flux of neutrinos from the whole population of clusters, as measured at Earth, integrated over the entire redshift range within the Hubble time (solid brown curve in the panels). In the left panel of Fig.~\ref{fig:Neu_FangSimM} and in Fig.~\ref{fig:Neu_FangSimz}, the injected CR spectrum is assumed to follow $E^{-1.5}$, with an exponential cut-off $E_\text{max} = 5\times 10^{16} \; \text{eV}$. Also, we  assumed in these cases that  $0.5\%$  of the kinetic energy of the clusters is converted to the CRs. Besides the total flux, this panel also shows the flux of neutrinos for several cluster mass intervals. 
The softening effect at higher energies is  due to the shorter diffusion time of the CRs, and to  the mass distribution of the clusters, as higher flux reflects lower population of massive clusters.  In Fig.~\ref{fig:Neu_FangSimz} we present  the integrated flux in  different redshift intervals and it can also be seen that the clusters at high redshift contribute less to the total flux of neutrinos. Those  at $z>1$ barely contribute to the flux due to the low population of massive clusters and their large distances.
Fig.~\ref{fig:Neu_FangSimM} and~\ref{fig:Neu_FangSimz}  also compares our results with the IceCube observations. We see that for  the assumed scenario for CRs injection in left panel of Fig.~\ref{fig:Neu_FangSimM} and in Fig.~\ref{fig:Neu_FangSimz},
they can reproduce the  IceCube observations for  $E > 20 $ TeV. 
In right panel of Fig.~\ref{fig:Neu_FangSimM}, instead, we have assumed that  $2 \;\%$  of the kinetic energy of the clusters is converted into  CRs, with a CR energy power-law spectrum   $E^{-2}$, with  $E_\text{max}$ following  the dependence below with the cluster mass and magnetic field:
\begin{equation}\label{eq:emax}
E_\text{max} = 2.8 \times 10^{18} \left(\dfrac{M_\text{cluster}}{{10^{15} M_{\odot}}}\right)^{2/3} \left( \dfrac{B_\text{cluster}~\text{G}}{10^{-6}~ \text{G}} \right)~ \text{eV},
\end{equation}
which is 
similar to \citet{fang2016high}.  In this scenario we find that the clusters contribution to the neutrino flux is smaller than IceCube measurements.

For all diagrams of  Fig.~\ref{fig:Neu_FangSimM} \& \ref{fig:Neu_FangSimz}, we also compare our results with those of \citet{fang2016high}) (blue lines). The total fluxes in both are similar, in general.Moreover, we see that in both cases, the largest contribution to the flux of  neutrinos comes from the cluster mass group $10^{14}~M_{\odot} < M < 10^{15}~M_{\odot}$.
However,  the contribution  from the mass group $10^{12}~M_{\odot} < M < 10^{14}~M_{\odot}$  in our results is a factor twice larger than that of \citet{fang2016high}, and smaller  by the same factor for the mass group $M > 10^{15}~M_{\odot}$, at  energies $E > 0.01$~PeV (left panel of Fig.~\ref{fig:Neu_FangSimM}).

A striking difference between the two results is that, according to \citet{fang2016high}, the redshift range  $0.3 \leq z \leq 1$ amounts for the largest contribution to neutrino production, but in our case the  redshift range $0.01 \leq z \leq 0.3$ provides a more significant contribution (see Fig.~\ref{fig:Neu_FangSimz}).
Besides, there is a difference of factor $\sim 2$ to  $\sim 3$ between ours and their results at these redshift ranges. 
This difference may be due to the more simplified modeling of the background distribution of clusters
in their case specially for the lower mass group ($10^{12}~M_{\odot} < M < 10^{14}~M_{\odot}$) at  high redshifts ($z > 1$).

\begin{figure}
\centering
\includegraphics[width=\sizeFigL]{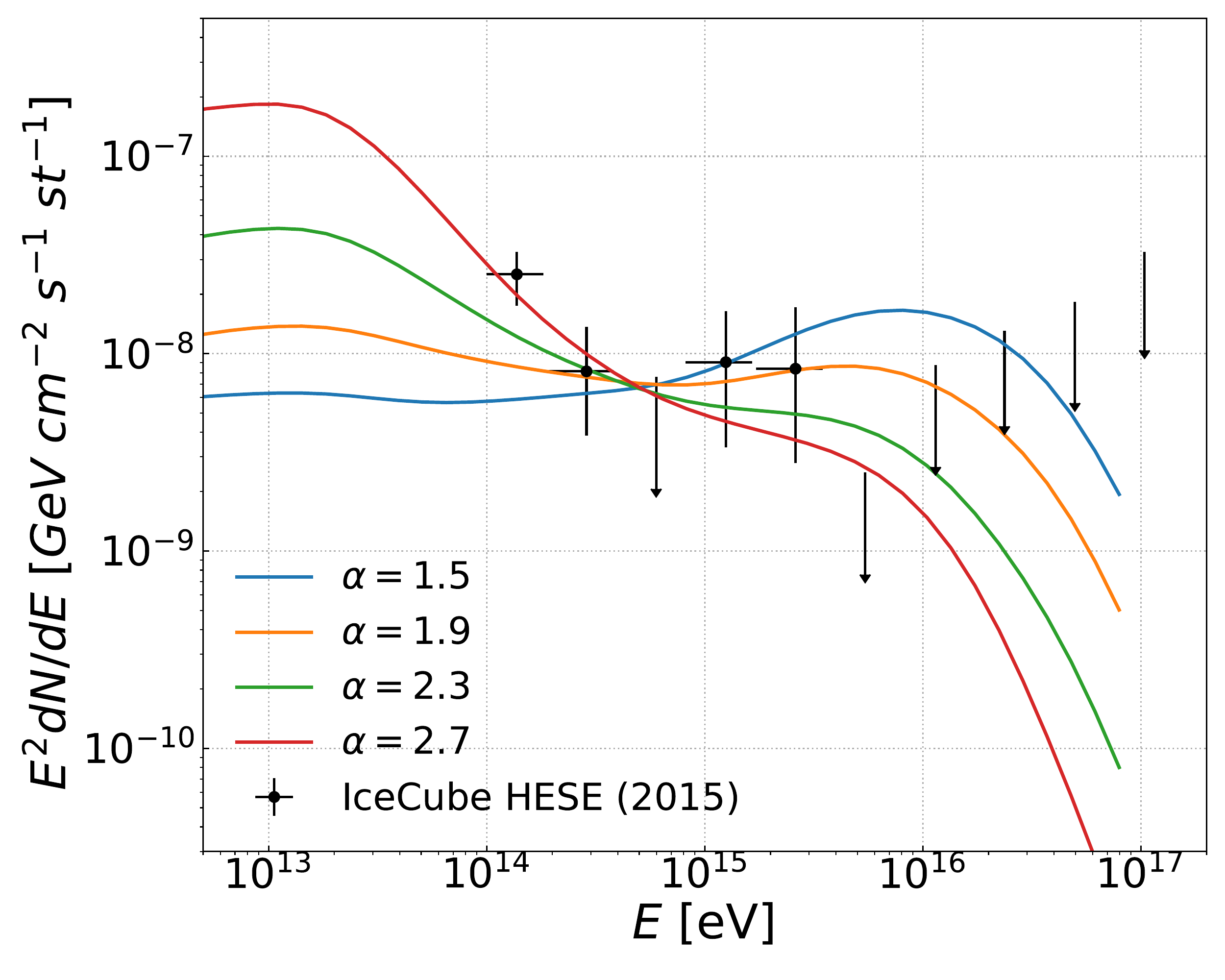}
\caption{Total spectrum of neutrinos for different injected CR spectra, $\sim E^{-\alpha}$, with  $\alpha = 1.5$ (blue), $1.9$ (orange), $2.3$ (green), $2.7$ (red). We consider $E_\text{max} = 5\times 10^{17}~\text{eV}$. 
This figure does not include the redshift evolution of the CR sources, $\psi_{ev}=1$ in  equation
equation (\ref{eq:flux_calc}). 
}
\label{fig:NeuSp_alpha}
\end{figure}

\begin{figure}
\centering
\includegraphics[width=\sizeFigL]{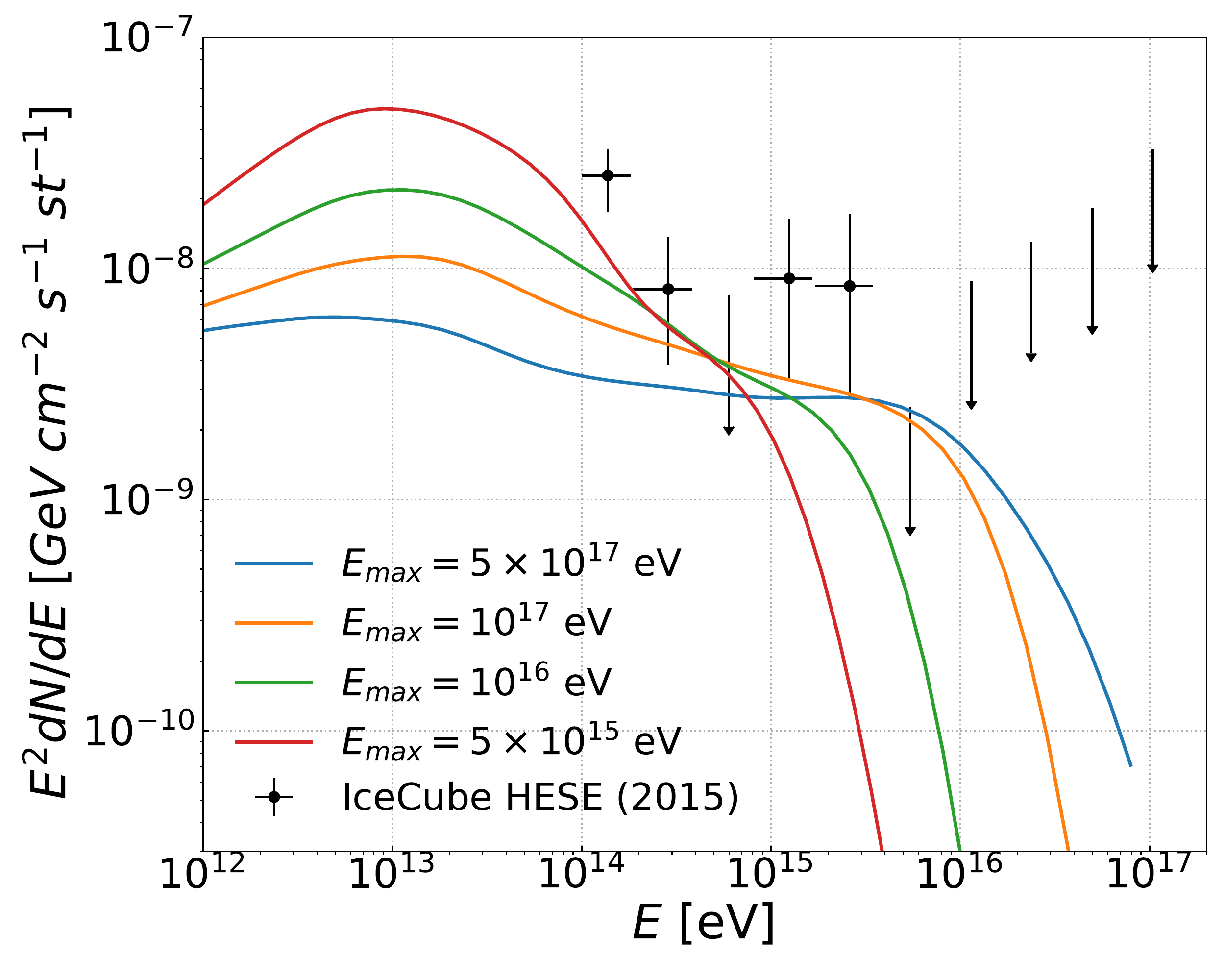}
\includegraphics[width=\sizeFigL]{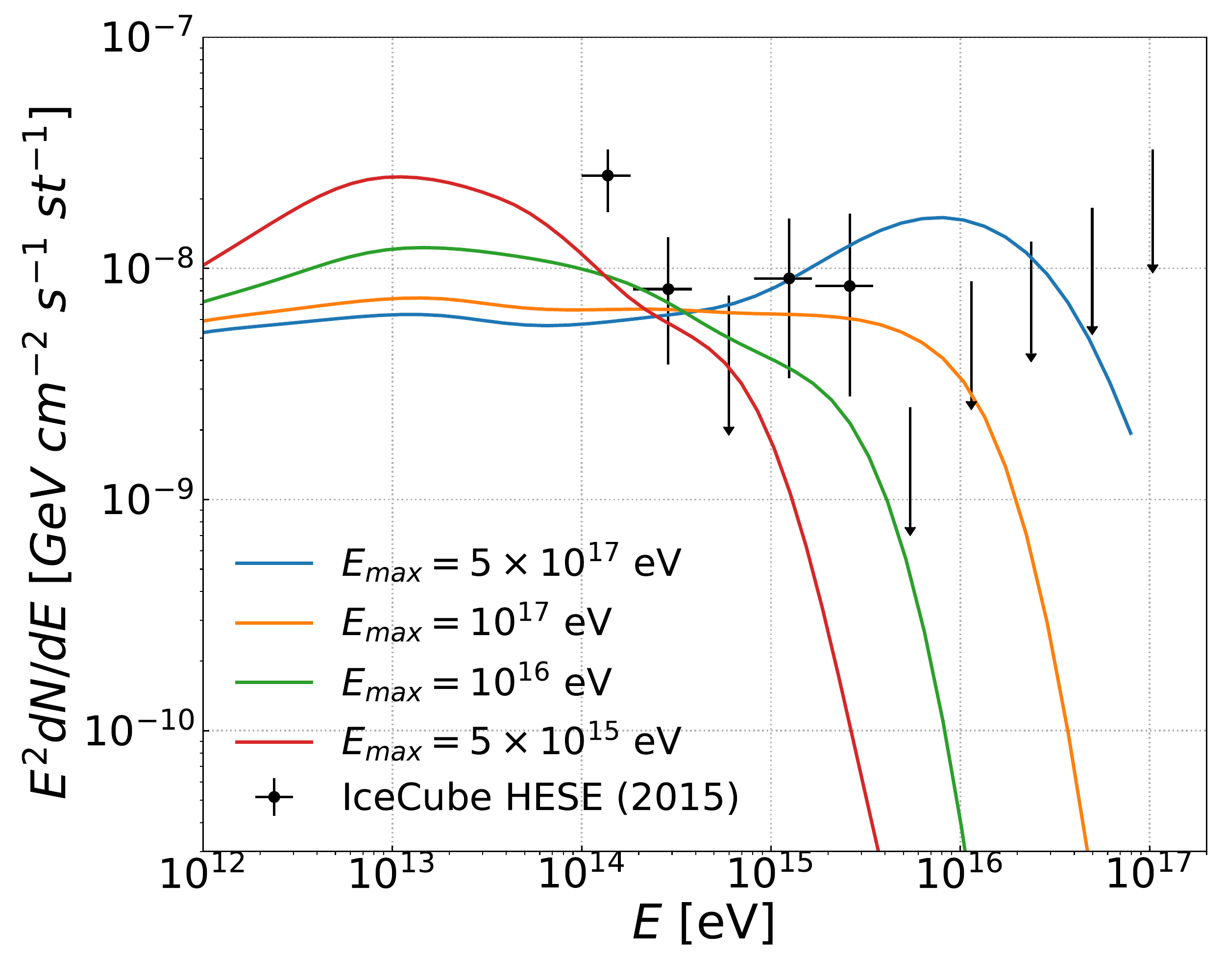}
\caption{Total neutrino spectrum for different cutoff energies i.e., $E_{max} = 5\times10^{15}$ (red), $10^{16}$ (green), $10^{17}$ (orange), and $5\times 10^{17}~ \text{eV}$ (blue). In the upper panel the spectral index is $\alpha =2$, and in lower panel  $\alpha =1.5$.
This figure does not include the redshift evolution of the CR sources, $\psi_{ev}=1$ in  equation
equation (\ref{eq:flux_calc}). 
}
\label{fig:Neu_Emax}
\end{figure}

In Fig.~\ref{fig:NeuSp_alpha}, we present the total neutrino spectra calculated for different spectral indices of the injected CRs, while in Fig.~\ref{fig:Neu_Emax} we show the  total neutrino spectra calculated for several cut-off energies. In order to try to fit the observed IceCube data, we have considered a $3 \; \%$ conversion of the kinetic energy of the cluster into CRs  in  Figs.~\ref{fig:NeuSp_alpha} \& \ref{fig:Neu_Emax}.

\begin{figure}
\centering
\includegraphics[width=\sizeFigL]{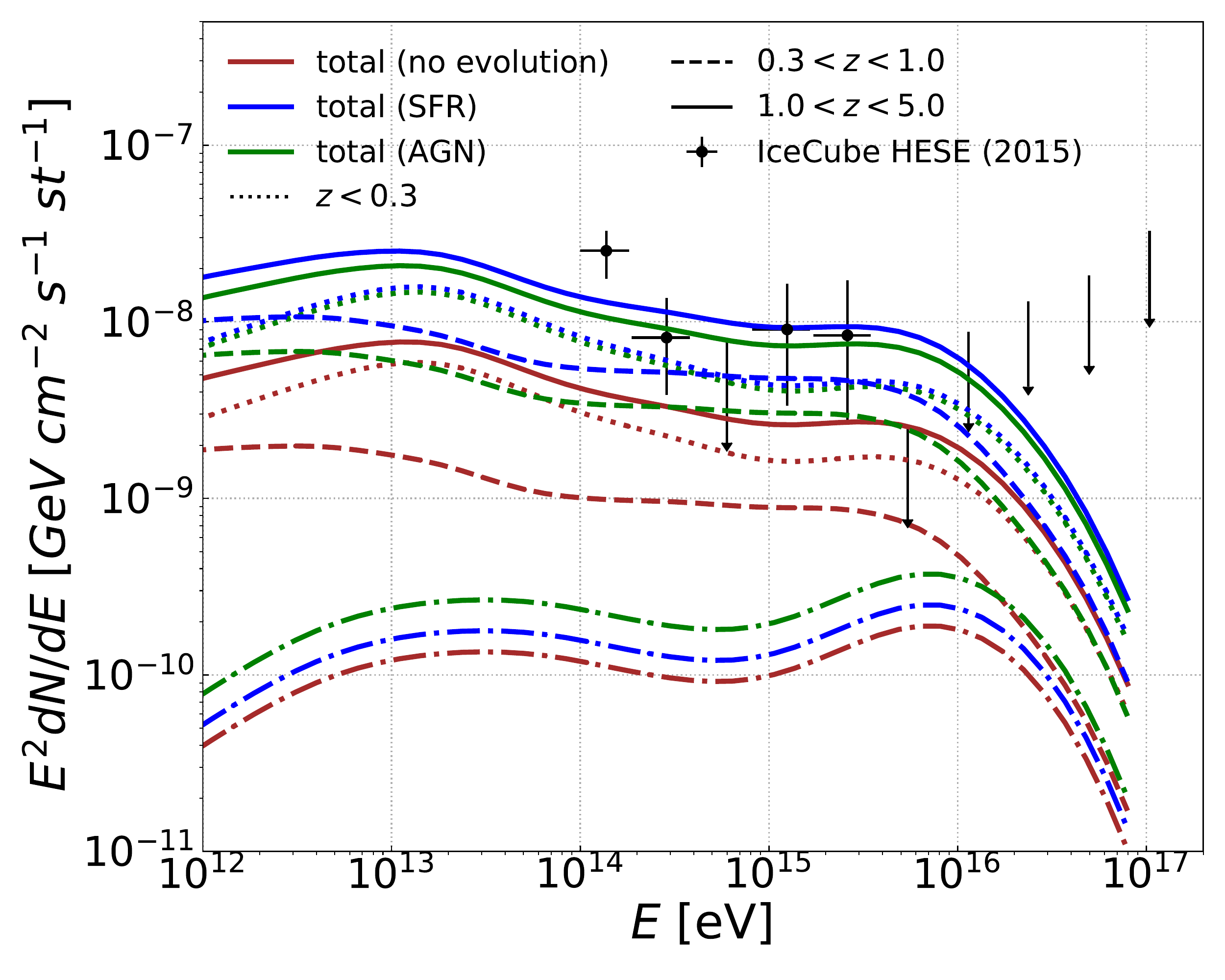}
\caption{Neutrino spectrum for different assumptions on the evolution of the CR sources: SFR (blue), AGN (green), and no evolution (brown). The fluxes are shown for different redshift ranges: $z < 0.3$ (dotted lines), $0.3 < z < 1.0$ (dashed), and $1.0 < z < 5.0$ (dash-dotted lines).  The CR injection spectrum has parameters $\alpha=2$ and $E_\text{max} = 5 \times 10^{17} \; \text{eV}$.}
\label{fig:NeuSP_ev}
\end{figure}

%\textcolor{red}
So far, we have computed the CR and neutrino fluxes from the clusters, considering no evolution function with redshift for both CR sources, AGN and SFR, i.e. we assumed $\psi_{\text{ev}}(z)=1$ in equation (\ref{eq:flux_calc}).  In Fig.~\ref{fig:NeuSP_ev},  we have included these contributions and plotted the flux of neutrinos for the redshift ranges: $z < 0.3, \; 0.3 < z < 1.0$, and $1.0 < z < 5.0$.
The flux is obtained for spectral index $\alpha = 2$ and cutoff energy $E_\text{max} = 5\times 10^{17} \; \text{eV}$.

%\textcolor{orange}
Clusters can directly accelerate CRs through shocks, but any type of astrophysical object that can produce HECRs can also contribute to the diffuse neutrino flux. In the former case, the sources evolve only according to the background MHD simulations, dubbed here ``no evolution'', whereas in the latter some assumptions have to be made regarding the CR sources.
In Fig.~\ref{fig:NeuSP_ev} we illustrate the impact of the source evolution.
We consider, in addition to the case wherein sources do not evolve, SFR and AGN-like evolutions (see equations~\ref{eq:AGNz} and \ref{eq:SFRz} and accompanying discussion). 
Our results suggest that, while the neutrino fluxes for the AGN and the SFR evolutions are relatively close to each other, the case without evolution contributes slightly less to the total flux. Moreover, at high redshifts ($1.0 < z <5.0$), AGNs in clusters produce more neutrinos than sources with SFR-like evolutions, whereas the same is not true for $z \lesssim 1$.

\begin{figure}
\centering
\includegraphics[width=\sizeFigL]{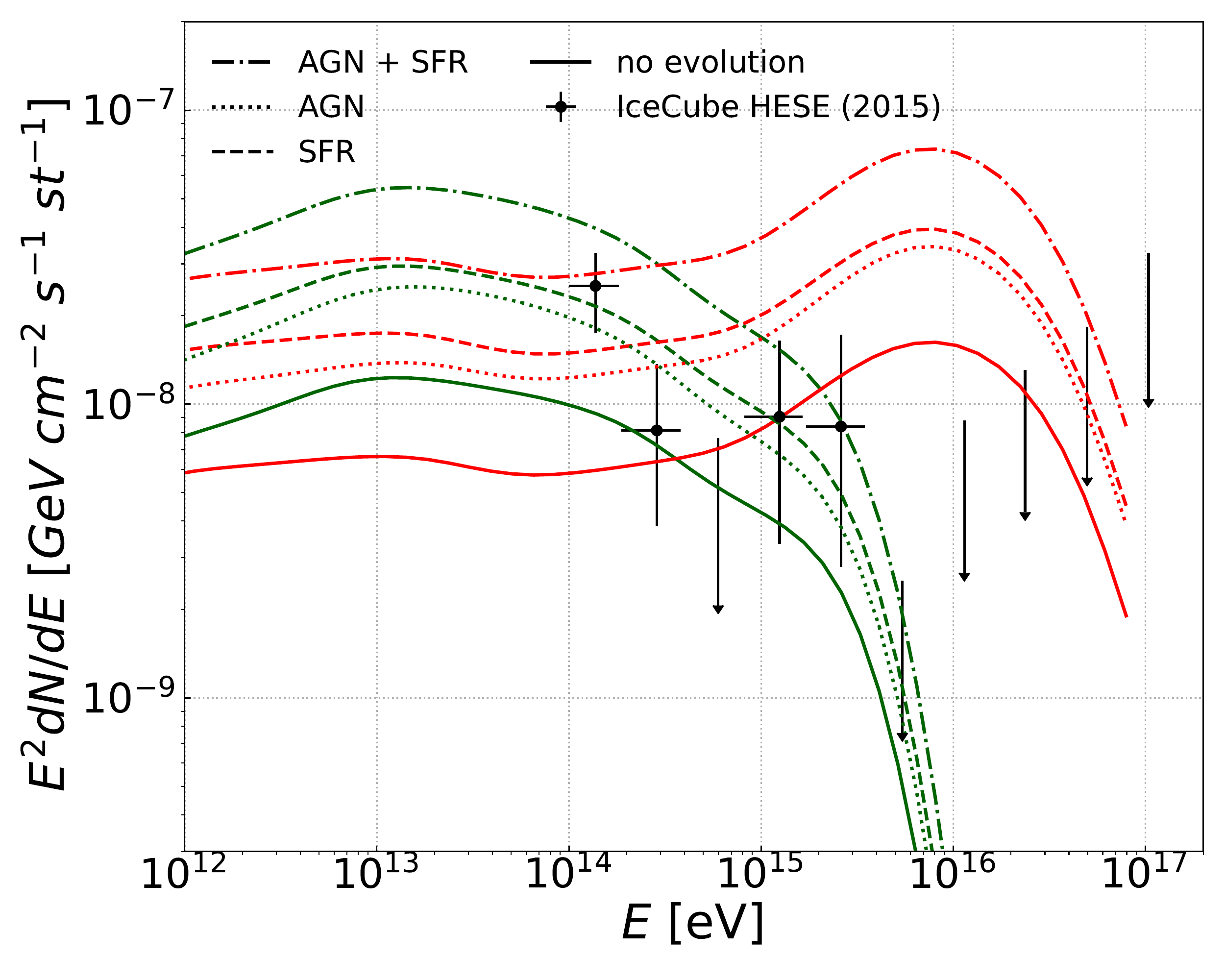}
\includegraphics[width=\sizeFigL]{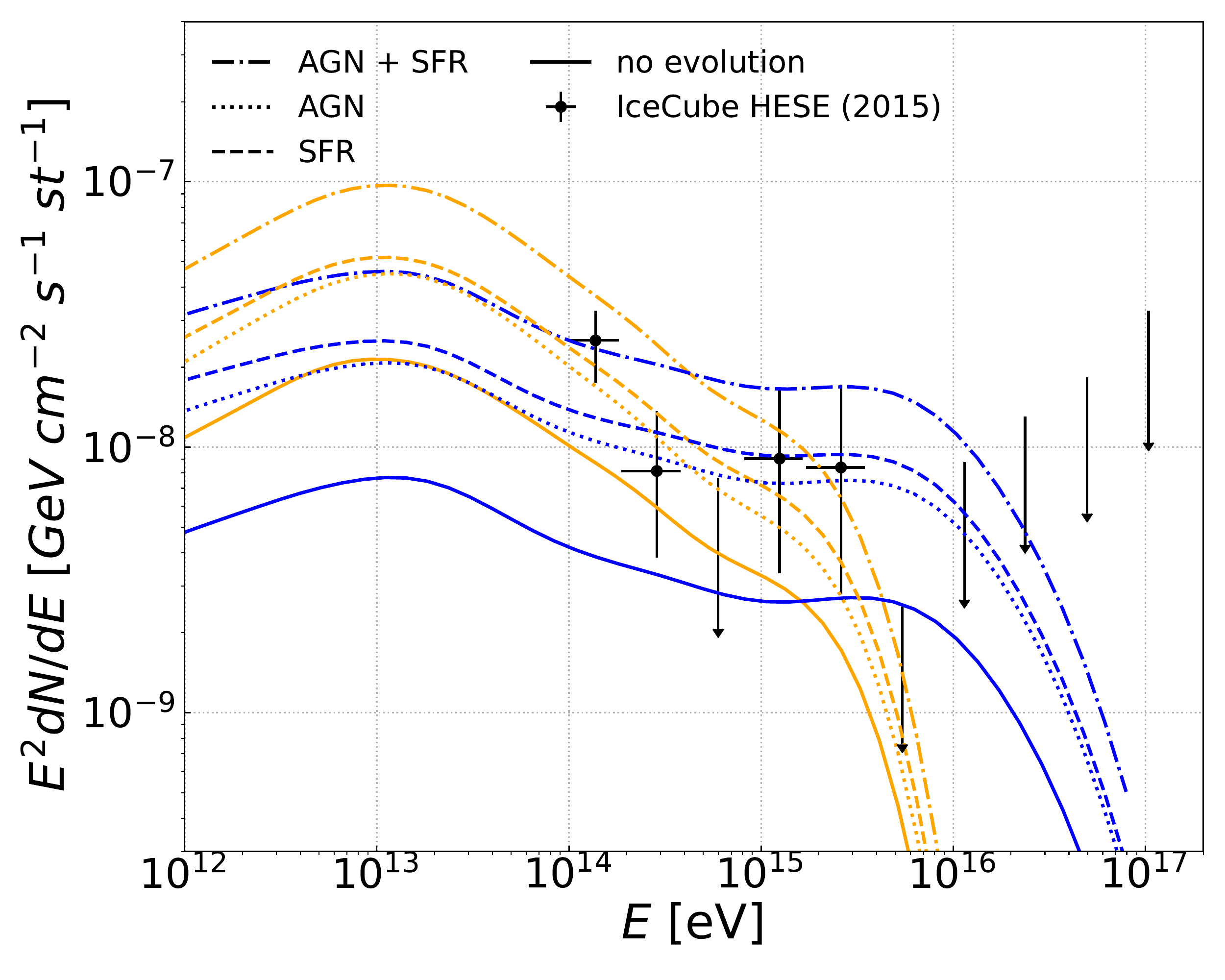}
\caption{Flux of neutrinos for different assumptions on the evolution of the CR sources: no evolution (solid lines), SFR (dashed lines), AGN (dotted lines) and AGN $+$ SFR (dash-dotted lines).
In upper panel green and red lines represent $\alpha=1.5$ for $E_\text{max} = 10^{16}$ and $5\times 10^{17} \; \text{eV}$ respectively. In lower panel orange and blue lines correspond to $\alpha=2$ for $E_\text{max} = 10^{16}$ and $5\times 10^{17} \; \text{eV}$, respectively.  }
\label{fig:NeuSP_AGN_SFR}
\end{figure}

%\textcolor{red}
In Fig.~\ref{fig:NeuSP_AGN_SFR}, we plotted the flux for different combinations of spectral index $\alpha$ and  $E_{max}$, with different source evolution assumptions as in Fig. \ref{fig:NeuSP_ev}.
In both panels all the combinations of $\alpha$ and $E_\text{max}$  are roughly matching with IceCube data, except  $\alpha = 1.5$, and  $E_{max} = 5\times 10^{17} \; \text{eV}$ in the upper panel as it overshoots the IceCube points.

%%%%%%%%%%%%%%%%%%%%%%%%%%%%%%%%%%%%%%%%%%%%%%%%%%%%%%%%%%%%%%%%%%%%%%%%%%%%%%%%%%%%
%%%%%%%%%%%%%%%%%%%%%%%%%%%%%%%%%%%%%%%%%%%%%%%%%%%%%%%%%%%%%%%%%%%%%%%%%%%%%%%%%%%%

\section{Discussion}\label{sec:Discussion}

In our simulations, the central magnetic field strength and gas number density  of the ICM are $\sim 10~ \mu \text{G}$ and $\sim 10^{-2}~ \text{cm}^{-3}$, respectively, for a cluster with mass $10^{15}~M_{\odot}$ at $z=0.01$,  and both  decrease toward the outskirts of the cluster. These quantities depend on the mass of the clusters, being smaller for less massive clusters (see Fig.~\ref{fig:Cluster-contour} \&~\ref{fig:ClusterB}). Thus, high-energy CRs will escape with a higher probability without much interactions in the case of less massive clusters. Lower-energy CRs, on the other hand, contribute less to the production of high-energy neutrinos. Therefore, we have a lower neutrino flux from  less massive clusters. In contrast, for massive clusters, higher magnetic field and gas density produce higher neutrino flux due to the longer confinement time, as we see in Fig.~\ref{fig:Neu-M}.

We tested several injection CRs spectral indices ($\alpha \simeq 1.5 - 2.7$), cut-off energies  ($E_{\text{max}} = 5\times10^{15} - 10^{18}$ eV), and source evolution (AGN, SFR, no evolution), in order to try to interpret the IceCube data  (see Figs.~\ref{fig:Neu_FangSimM},~\ref{fig:Neu_FangSimz},
~\ref{fig:NeuSp_alpha},~\ref{fig:Neu_Emax},~\ref{fig:NeuSP_ev} and \ref{fig:NeuSP_AGN_SFR}). 
Overall, our results indicate that  galaxy clusters can contribute to a considerable fraction of the diffuse neutrino flux measured by IceCube at energies between $100 \; \text{TeV}$ and $10 \; \text{PeV}$, or even all of it, provided that that protons compose most of the CRs.

Our results also look, in principle, similar to those of \citet{fang2016high} with no source evolution, who considered essentially the same redshift interval, but employed  semi-analytical profiles to describe the cluster properties. In particular, in both cases, the largest contribution to the flux of  neutrinos comes from the cluster mass group $10^{14} < M < 10^{15}~M_{\odot}$. However, they did not consider the interactions of CRs with CMB and EBL background as they considered it subdominant compared to the hadronic background  following  \citet{kotera2009propagation}. But, it can be seen from the upper panel of Fig.~\ref{fig:MFP_Neu-Traj} that $\lambda$ for pp-interaction and photopion production in the CMB are comparable for CRs of energy $\gtrsim 10^{17}$~eV. Therefore, the neutrino production due to CR interactions with the CMB  is not negligible. Perhaps the most relevant difference between our results and theirs  is that, in their case, the redshift range  $0.3 \leq z \leq 1$ makes the largest contribution to neutrino production, while in our case this comes from the redshift range $z \lesssim 0.3$, when considering no source  evolution (see Fig~\ref{fig:Neu_FangSimz}).

%\textcolor{magenta}
When including source evolution, 
there is also a dominance in the neutrino flux from the redshift range $z \lesssim 0.3$, though the contribution due to the evolution of  star forming galaxies (SFR) from redshifts $0.3 \leq z \leq 1$ is also important. 
Overall, the inclusion of source evolution can increase the diffuse neutrino flux by a factor of  $\sim 3$ (when considering the separate contributions of AGN or SFR) to  $\sim 5$ (when considering both contributions concomitantly) in the cases we studied, compared to the case with no evolution,  which is in agreement with \citep{murase2016constraining}.
Also, our results agree with the IceCube measurements for $E \gtrsim 10^{14} \; \text{eV}$ and are in rough accordance with \citep{murase2017active, fang2018linking}. 
Nevertheless, since there are uncertainties related to the choice of  specific populations for the CR sources, obtaining a full picture of the diffuse high-energy neutrino emission by clusters is not a straightforward task.

It is also worth comparing our results with \citet{zandanel2015high}, who evaluated the neutrino spectrum based on  estimations of the radio to gamma-ray luminosities of the clusters in the universe. Although our work has assumed  an entirely different approach, 
both results are consistent,  especially for a CR spectral index $\alpha \simeq 2$. 
High-energy ($E>10^{17}$~eV) CRs can escape easily from clusters, 
effectively leading to a spectral steepening that was not considered by \citet{zandanel2015high}. However, not all the clusters are expected to produce hadronic emission \citep{zandanel2015high,zandanel2014physics}. In fact, we observe less hadronic interactions in the case of low-mass clusters ($M \lesssim 10^{14} \; M_{\odot}$), which could further limit the neutrino contribution from clusters.

The cluster scenario may  get strong backing due to anisotropy detections above PeV energies. Recently, only a few sources  of high-energy neutrinos have been observed \citep{aartsen2013first, aartsen2015atmospheric, albert2018search,ansoldi2018blazar, aartsen2020time}, 
but there are also expectations to increase the observations  with future instruments like IceCube-Gen2~\citep{icecube2020gen2}, KM3NeT~\citep{km3net2016}, and the Giant Radio Array for Neutrino Detection (GRAND) \citep{alvarez2020giant}. Specifically, neutrinos from clusters are more likely to be observed if the flux of cosmogenic neutrinos is low, which might contaminate the signal, as discussed by \citet{batista2019cosmogenic}.

%%%%%%%%%%%%%%%%%%%%%%%%%%%%%%%%%%%%%%%%%%%%%%%%%%%%%%%%%%%%%%%%%%%%%%%%%%%%%%%%%%%%
%%%%%%%%%%%%%%%%%%%%%%%%%%%%%%%%%%%%%%%%%%%%%%%%%%%%%%%%%%%%%%%%%%%%%%%%%%%%%%%%%%%%
\section{Conclusions}\label{sec:conclusion}

We  considered a cosmological background based on  3D-MHD simulations  to model the cluster population of the entire universe,  and a multidimensional Monte Carlo technique to study the propagation of CRs in this environment and obtain the flux of neutrinos they produce. 
Our results can be summarized as follows:
\begin{itemize}

\item
We found that  CRs with energy $E\lesssim 10^{17}$ eV cannot escape from the innermost regions of the clusters, due to interactions with the background gas, thermal photons and magnetic fields. Massive clusters 
($M \gtrsim 10^{14} \; M_{\odot}$) have stronger magnetic fields which can confine these high-energy CRs for a time comparable to the age of the universe.

\item
Our simulations predict that the neutrino flux above PeV energies comes from the most massive clusters because the CR interactions with the gas of the ICM are rare for clusters with $M < 10^{14} \; M_{\odot}$. 

\item
Most of the neutrino flux comes  from nearby clusters in the redshift range $z \lesssim 0.3$. The high-redshif clusters contribute less to the total flux of neutrinos compared to the low-redshift ones, as the population of massive clusters at high redshifts is low.

\item
The total integrated neutrino flux obtained from the interactions of CRs with the ICM gas and CMB during their propagation in the turbulent magnetic field can account for sizeable percentage of the IceCube observations, especially, between energy $100 \; \text{TeV}$ and $10 \; \text{PeV}$.

\item 
Our results also indicate that the redshift evolution of  CR sources like AGN and SFR, enhance the flux of  neutrinos.

\end{itemize}

Finally, more realistic studies  considering  cosmological simulations that account for AGN and star formation feedback from galaxies \citep[ e.g. ][]{barai_etal2016, barai_dalpino2019}
will allow to constrain better the redshift evolution of the CR sources in the computation of the total neutrino flux from clusters. Furthermore, in the future, IceCube will have detected more events. Then, combined with diffuse gamma-ray searches by the forthcoming CTA~\citep{cta2019}, it will be possible to better assess the contribution of galaxy clusters to the total extragalactic neutrino flux.

\section*{Acknowledgements}

Saqib Hussain acknowledges support from the Brazilian funding agency CNPq.   EMdGDP is also grateful for the support of the Brazilian agencies FAPESP (grant 2013/10559-5) and CNPq (grant 308643/2017-8). RAB is currently funded by the Radboud Excellence Initiative, and received support from FAPESP in the early stages of this work (grant 17/12828-4). KD  acknowledges support by the Deutsche Forschungsgemeinschaft (DFG, German Research Foundation) under Germany’s Excellence Strategy – EXC-2094 – 39078331 and by the funding for the COMPLEX project from the European Research Council (ERC) under the European Union’s Horizon 2020 research and innovation program grant agreement ERC-2019-AdG 860744.
The numerical simulations presented here were performed in the cluster of the Group of Plasmas and High-Energy Astrophysics (GAPAE), acquired with support from FAPESP (grant 2013/10559-5). This work also made use of the computing facilities of the Laboratory of Astroinformatics (IAG/USP, NAT/Unicsul), whose purchase was also made possible by a  FAPESP (grant 2009/54006-4). 
We also acknowledge  very useful comments from K. Murase on an earlier version of this manuscript.

\bibliographystyle{mnras}
\bibliography{ICM_Neu_MNARS}

\appendix

\section{Mean Free Paths }\label{Appsec:MFP}

The mean free path $\lambda$ for  different CR interactions in the ICM are defined below.

For a CR proton with Lorentz factor $\gamma_p$ traversing an isotropic photon field, one obtains the rate $\lambda_{p\gamma}^{-1}(E_p)$ \citep{schlickeiser2002interplanetary}
\begin{align}
\lambda_{p\gamma}^{-1}(E_p)  &= \frac{1}{2\gamma_{p}}\int\limits_{\epsilon_{\text{th}}/2\gamma_p}^{\infty}d\epsilon \frac{n_\text{ph}(\epsilon, r_i)}{\epsilon^2}\int\limits_{\epsilon_{\text{th}}}^{2\gamma_p \epsilon} d\epsilon^{\prime} \epsilon^{\prime}\sigma_{p\gamma}(\epsilon^{\prime}) K_{p}(\epsilon^{\prime}), \label{eq:MFPpg} 
\\
\epsilon_{\text{th}} &= K m_{\pi}c^2\left[1+\frac{K m_{\pi}}{2m_p}\right] = 145 \; \text{MeV}.
 \label{eq:thresholdE}
\end{align}
Where  $n_\text{ph}(\epsilon, r_i)$ denotes the number density of photons of energy $\epsilon$ at a given distance $r_i$ from the center of the cluster and $\sigma_{p\gamma}$ is the cross section of  the interaction of CRs with background photons.
The threshold energy for the production of $K$ pions is given by equation (\ref{eq:thresholdE}), so that for the production of a single ($K=1$)
pion the rest system threshold energy is $\epsilon_{\text{th}} = 145~ \text{MeV}$ \citep{schlickeiser2002interplanetary}.

To calculate the rate for the interactions of high-energy photons (produced during the propagation of CRs inside a cluster) with the local protons in the ICM, we can use  
equation (\ref{eq:MFPpg}) with the following modification in the center-of-mass (CM) energy.
The energy $E$ and 3-momentum ${\bf p}$ of a particle of mass $m$ form a 4-vector $p = (E, p)$
whose square $p^2 = (E/c)^2 - {\bf p}^2 = m^2c^4$. 
The velocity of the particle is $\beta c = {\bf v}/c = {\bf p}/E$.
In the collision of two particles of masses $m_1$ and $m_2$, the total CM energy can be expressed in the Lorentz-invariant form as
\begin{align} 
\epsilon_{CM} &=  \left[\frac{1}{c^2}(E_1 + E_2)^2 - ({\bf p_1} + {\bf p_2} ) c^2\right]^{1/2} \\ 
&= \left[m_1^2c^4 + m_2^2c^4 + \frac{2E_1E_2}{c^2}(1-{\bf \beta_{1} \beta_{2}} \cos\theta) \right]^{1/2},
\end{align}
where $\theta$ is the angle between the particles that  we can consider zero. In the frame where one particle (of mass $m_2$)
is at rest (lab frame) then,
\begin{equation}
\epsilon_{CM} = (m_1^2 c^4 + m_2^2 c^4 + 2E_1 m_2 c^2)^{1/2}.
\end{equation}
If we consider $m_2$ is proton and $m_1$ is photon, then the above relation becomes
\begin{equation}\label{eq:E_cm}
\epsilon_{CM} = (m_2^2 c^4 + 2E_1 m_2 c^2)^{1/2}.
\end{equation}

\begin{equation}\label{eq:MFPGammap}
\lambda_{\gamma p}^{-1}(\epsilon_\text{ph})  = \frac{\epsilon_p}{2\epsilon_\text{ph}} \int\limits_{\epsilon_{\text{th}}/(2\epsilon_\text{ph}/\epsilon_p)}^{\infty}d\epsilon \frac{n_{p}(\epsilon, r_i)}{\epsilon_p^2} \int\limits_{\epsilon_{\text{th}}}^{(2\epsilon_\text{ph}/\epsilon_p) \epsilon} d\epsilon^{\prime} \epsilon^{\prime}\sigma_{\gamma p}(\epsilon^{\prime}),
\end{equation} 
so that the rest frame is in the
local protons.
We used equation (\ref{eq:E_cm}) for the energy of the CM in equation (\ref{eq:MFPGammap}). 
In equation (\ref{eq:MFPGammap}), $n_{p}(\epsilon, r_i)$  is number density of local protons with energy $\epsilon_p = m_p c^2 \sim 1$~GeV at a given distance $r_i$ from the center of a cluster and decreases toward the outskirt,  $\epsilon_{\text{th}} \sim 1.4 \times  10^{8}$~eV is the threshold energy for this interaction and the cross section $\sigma_{\gamma p}(\epsilon^{\prime})$ is of the order $ \sim 10^{-37}~(\text{cm}^{2})$. With these values used in equation (\ref{eq:MFPGammap}) we solve this integral to calculate  $\lambda$ for $\gamma$-proton interaction. 
We calculated  $\lambda$ 
from equations \ref{eq:MFPpg}-\ref{eq:MFPGammap} with some modifications to include the information of the spatially dependent Bremsstrahlung photon field of the clusters $n_\text{ph}(\epsilon, r)$.

For proton-proton (pp) interaction, the rate  is given by 
\begin{equation}\label{eq:ratepp}
\lambda^{-1}_{\text{pp}} (E_p, r_i) = K_{\text{pp}}~\sigma_{\text{pp}}(E_p)~ n_i (r_i) 
\end{equation}
Where $K_{\text{pp}} = 0.5$ is the inelasticity factor,
$n_i (r_i)$ denotes the number density of proton  at a given distance $r_i$ from the center of the cluster and $E_p$ is the energy of the protons.

To obtain the proton number density, we consider that the background plasma consists of electrons and protons in near balancing. Since the abundance is mostly of H and this is mostly ionized in the hot ICM, this is a reasonable assumption.  Thus  $n_p \simeq n_e$, and  $\rho_{\text{gas}} = n_p m_p + n_e m_e \sim n_p m_p$, so that  
$n_{i} \simeq n_e \simeq \rho_{\text{gas}} / m_{p}$,
where $m_p$ is the proton mass and $\rho_{\text{gas}}$ is the gas mass density in the system. 

For $\sigma_{\text{pp}} = 70~ \text{mb}$ ($1\text{barn} = 10^{-29} \; \text{m}^{2}$), we have for the cross section \citep{kafexhiu2014parametrization}:
\begin{equation}\label{eq:sigmapp}
\sigma_{\text{pp}} = \left[30.7 - 0.96\log\left(\frac{E_p}{E_{p}^{\text{th}}}\right) + 0.18\log\left(\frac{E_p}{E_{p}^{\text{th}}}\right)\right]  \left[1-\left(\frac{E_{p}^{\text{th}}}{E_p}\right)^{1.9}\right]^3~ \text{mb},
\end{equation}
where $E_p$ is the energy of the proton and $E_{p}^{\text{th}}$ is the threshold kinetic energy $E_{p}^{\text{th}} = 2m_{\pi}  + m_{\pi}/m_{p} \approx 0.2797$ GeV.
We used equations \ref{eq:ratepp} and (\ref{eq:sigmapp}) to calculate  $\lambda_{\text{pp}}$.

\section{Spectral Index}\label{Appsec:specIndex}
To calculate the flux of 
neutrinos corresponding to  injected  CRs with
an arbitrary power-law spectrum with power law index $\alpha$, 
$dN_{\text{CR},~ E}/dE \propto E_i^{-\alpha} \exp\{-E_i/E_\text{max}\}$,
we can normalize
the spectrum as follows:
\begin{equation}\label{eq:jalpha}
J(\alpha) = \dfrac{\ln(E_{\text{CR},~ \text{max}}/E_{\text{min}})}{\int\limits_{E_{\text{min}}}^{E_{\text{CR, max}}}E_i^{1-\alpha} \exp\left(-\frac{E_i}{E_\text{max}}\right)dE}  E_i^{1-\alpha} \exp\left(-\frac{E_i}{E_\text{max}}\right)
\end{equation}
Where, $E_{i}$ is the injection energy of the simulated CRs, 
$E_\text{max}$ is the exponential cut-off energy, and $E_{\text{CR, max}}$ is the maximum injection energy of the CRs.

\end{document}